\documentclass[11pt]{article} 
\usepackage{graphicx} 
\usepackage{amsmath}
\usepackage{amssymb} 
\usepackage{epstopdf}
\usepackage{esint}

\title{Quantum Stochastic Processes: A Case Study} \author{} \date{}

\begin{document}
\maketitle

\vskip -1.0 truecm

\centerline{\large Michel Bauer ${}^{\spadesuit,\clubsuit~}$\footnote{
\texttt{michel.bauer@cea.fr}} and Denis Bernard
${}^{\clubsuit~}$\footnote{Member of C.N.R.S; \texttt{denis.bernard@ens.fr}} }

\bigskip

\centerline{${}^\spadesuit$ \large Institut de Physique Th\'eorique de Saclay
\footnote{CEA/DSM/IPhT, Unit\'e de recherche associ\'ee au CNRS} ,}
\centerline{CEA-Saclay, 91191 Gif-sur-Yvette, France.}  \medskip
\centerline{$^\clubsuit$ \large Laboratoire de Physique Th\'eorique de l'Ecole
Normale Sup\'erieure,} \centerline{CNRS/ENS, Ecole Normale Sup\'erieure, 24 rue
Lhomond, 75005 Paris, France}

\newcommand{\bC}{\mathbb{C}} 
\newcommand{\bR}{\mathbb{R}}
\newcommand{\bZ}{\mathbb{Z}} 
\newcommand{\bN}{\mathbb{N}}
\newcommand{\bD}{\mathbb{D}} 
\newcommand{\bH}{\mathbb{H}}

\newcommand{\sP} {\mathcal{P}} 
\newcommand{\sF}
{\mathcal{F}} 
\newcommand{\sL}{\mathcal{L}} 
\newcommand{\sD}{\mathcal{D}}
\newcommand{\ud}{\mathrm{d}} 
\newcommand{\bn}{\mathbf{n}} 
\newcommand{\Order}{\mathcal{O}} 
\newcommand{\order}{o} 
\newcommand{\unit}{\mathbf{1}}
\newcommand{\bra}{\langle} 
\newcommand{\ket} {\rangle} 
\newcommand{\prob}{\mathbb{P}} 
\newcommand{\expect}{\mathbb{E}} 
\newcommand{\cconj}{\overline}
\newcommand{\im}{\Im \textrm{m }} 
\newcommand{\re}{\Re \textrm{e }}
\newcommand{\const}{\mathrm{const.}}  
\newcommand{\mbar}[1]{#1^{\dagger}}
\newcommand{\mtild}[1]{\tilde{#1}}

\newtheorem{theorem}{Theorem} 
\newtheorem{lemma}{Lemma}
\newtheorem{corollary}{Corollary} 
\newcommand{\proof}{\emph{Proof: } }
\newcommand{\QED}{$\square$}

\def\vev#1{\langle{#1}\right}

\vskip 1.0 truecm

\begin{abstract} 
We present a detailed study of a simple quantum stochastic process, the quantum phase space Brownian motion, which we obtain as the Markovian limit of a simple model of open quantum system. We show that this physical description of the process allows us to specify and to construct the dilation of the quantum dynamical maps, including conditional quantum expectations. The quantum phase space Brownian motion possesses many properties similar to that of the classical Brownian motion, notably its increments are independent and identically distributed. Possible applications to dissipative phenomena in the quantum Hall effect are suggested.
\end{abstract}

\newpage
\tableofcontents 
\newpage

\section{Introduction and motivations}
Quantum Brownian motion \cite{CL} is usually defined by coupling quantum mechanically
a massive particle to a bath of harmonic oscillators, see e.g.
refs.\cite{Gard-Zoller,Breuer-Pet,Legget-review} and references therein. 
This is not the quantum analogue of what probabilists call
Brownian motion (which is the continuum limit of random walks) but that of a
Newtonian particle submitted to friction and random force (which is also called
Brownian motion in the physics literature), whose dynamics is described by
$m \ddot{x} + \alpha \dot x = f$ with $\alpha$ the friction coefficient and $f$
the random forcing. These two notions are related: the former is the over-damped
limit, or the massless limit, of the latter. The mass plays a role of short
distance cut-off, and the velocity $\dot{x}$ scales as $1/\sqrt{m}$. It diverges
in the probabilistic Brownian motion which is known to be nowhere
differentiable. This causes troubles when defining the quantum analogue of the
probabilistic Brownian motion as the massless limit of the quantum Brownian
motion because the momentum $p=m\dot{x}$ scales as $p\sim \sqrt{m}$, whose
massless limit is incompatible with the canonical commutation relation
$[x,p]=i\hbar$. The first aim of the present paper is to understand how to
consistently formulate the over-damped quantum Brownian motion.

A possible way to overcome these difficulties may consist in going one dimension
higher and consider the two dimensional Brownian motion. An advantage is that
one may then couple the Newtonian particle to an out-of-plane external magnetic
field $B$, so that the classical equation of motion becomes
\[ m\ddot{z} + (\alpha +i eB)\dot{z} = f \] 
in complex coordinate $z=x+iy$, with $e$ the electric charge. If the force $f$
is white-noice in time, i.e. $\langle f(t)\bar f(s)\rangle\propto \delta(t-s)$,
the massless limit still describes the probabilistic 2D Brownian motion, and its
quantification will provide a candidat for the 2D (over-damped) quantum Brownian
motion. The magnetic field only rotates and dilates the Brownian trajectories.
Since these trajectories are known to be statistically conformally invariant
\cite{levy}, the effect of the magnetic field is classically quite
innocent. Quantum mechanically the situation is different. In absence of random
forcing, the energy spectrum is that of the Landau levels. The magnetic length
$\ell_B=\sqrt{\hbar/eB}$ provides a short-distance cut-off which would allow us
to take the massless limit. Since the cyclotron frequency $\omega_B=eB/m$
diverges as $m$ vanishes (or as $B$ increases), taking the massless (or the over-damped) limit 
amounts to project on the first Landau level. 
Thus a model candidate for the over-damped 2D quantum Brownian
motion is a particle confined to the first Landau level and coupled to a 
harmonic oscillator bath \footnote{One may wonder whether this model may have
  relevance to the description of dissipative effects in the integer quantum
  Hall effect, although it does not deal with the edge currents.}. 
  Recall that, once projected onto the first Landau level, the
two coordinate operators $x$ and $y$ are not commuting \cite{landau} but form a pair of
canonical operators, $[x,y]=i\ell_B^2$, so that the motion actually
takes place in the quantum phase plane. The model obtained by following this
projection strategy leads to a coupling between the 
oscillator bath and the particle which (although quadratic) is yet too
complicated \footnote{We nevertheless hope to report on this model in the near
  future \cite{hope to come}.}. Hoping, or claiming, for universality property,
we consider an alternative model with a much simpler coupling between the
quantum phase plane and a bath of harmonic oscillators, see eq.(\ref{hmodel}). 

In this way, in a weak-coupling/long-time/continuous-density limit, we obtain a Markovian quantum process, which we call "quantum phase Brownian motion" and which bares many similarities with classical Brownian motion.
A similar, but not identical, model has been considered by E. Davies in ref.\cite{Davies73} 
but without analysing the quantum process.

The two dimensional Brownian motion possesses very peculiar geometrical
properties linked to conformal invariance, see e.g. ref.\cite{lawler}. It would
be very interesting to know how these properties are deformed after
quantisation. To understand the quantum geometry of the quantum 2D Brownian motion is
another long term motivation of the present work \cite{hope to come}.

The second aim of the paper is to provide a description -- as complete as
possible -- of the quantum stochastic process defined by this simple model. We
aim at presenting the algebraic structure underlying the process which goes
beyond the dynamical maps and the associated master equation, which code for the
evolution of the reduced density matrix \cite{Barch,Biane1,Attal}. This is in
particular needed if one is willing to evaluate, or even define, multi-point
expectations. Recall that a classical stochastic process is defined over a
probability space equipped with a probability measure and a filtration (i.e. an
increasing family of $\sigma$-algebras) to which the process is adapted. The
process is specified by a consistent set of finite-dimensional distributions
(encoded in transition kernels for Markov processes) and
the filtration allows to define conditional expectations. The filtration encodes
the increase of knowledge as time and the process go on. In Appendix
\ref{App:stochastic} we recall the algebraic structure 
 induced by these data.  Quantum mechanically, transition
kernels are replaced by completely positive dynamical maps acting on the algebra
of observables of the quantum system (but not of the bath). By duality, this
specifies the evolution of the system density matrix and it is enough for
evaluating one-point functions (but not multi-point). Most, but not all, of the
physics literature only deals with density matrices. This is however not enough
to fully specify the process and in particular to define the quantum analogue
of conditional expectations. This requires extending the algebra of observables
and the dynamical maps to a set of embedded algebras (quantum analogues of the
filtration, which roughly keeps track of the process up to time $t$) with flows intertwining them and projectors (quantum analogues of
conditional expectations) with compatibility properties. This extended algebraic
structure, which mimics that of the classical processes (see Appendix
\ref{App:stochastic}), is called a dilation of the dynamical maps
\cite{Barch,Biane1,Attal,math:dilation}. Given completely positive maps, there is
generically not a unique dilation but a set of compatible dilations. We present
this structure in the simple (but quite generic) model we describe and show that
the physical description of the process given by coupling the quantum system to
an oscillator bath specifies the dilation \footnote{We show this property in the
  case of this simple quadratic model, but we hope to also proof it in a future
  publication \cite{hope to come} in the case of an arbitrary finite dimensional
  quantum system linearly coupled to a bath.}.

Exemple of dilations of dynamical maps have of course been given in the
literature, see e.g. refs.\cite{Barch,Biane1,Attal} and references therein. Elements
of the construction we present here also bare some similarities with the quantum
stochastic calculus of ref.\cite{Huds-parth}. However,
as far as we know, our construction of the dilation from the microscopic
physical model has not been described earlier. Our model provides a
simple framework in which concepts of quantum stochastic processes may be
explained and exemplified.

The paper is organised as follows: In the section \ref{Sec:bath}, we define the
model and describe the long-time/weak-coupling limit in which it becomes
Markovian. In section \ref{Sec:brown} we describe the quantum process with a
detailed presentation of the quantum filtration, of the dynamical maps and the
associated Lindbladian, of the flows and the conditional expectations, etc...
The quantum process we have defined enjoys many properties analogous to those of
Brownian motion. In particular we show that its increments are independent
and identically distributed and we show that the conditional expectations are
neutral with respect to both left and right mesurable
multiplications\footnote{This is explained precisely below.}. We also make
contact with processes introduced earlier on the mathematical literature. In
Appendix \ref{App:stochastic} we recall the algebraic structure underlying
classical stochastic processes and in Appendix \ref{App:weak} we present an
alternative approach to the weak-coupling/long-time limit.

\section{A simple heat bath model} \label{Sec:bath}

\subsection{The model} Out-of-equilibrium quantum systems may often be described
by coupling them to reservoirs which, quantum mechanically, may be modelled as
collections of harmonic oscillators. Here we consider one of the simplest model:
a harmonic oscillator (which we call the system oscillator) linearly coupled to
a reservoir set of independent harmonic oscillators (which we call the bath). We
denote by $\epsilon$ the energy of the system oscillator and by $\omega_\alpha$
the pulsations of the reservoir oscillators. The system creation-annihilation
operators will be denoted $\mbar w$ and $w$ and those of the reservoir by
$\mbar a_\alpha$ and $a_\alpha$. They satisfy canonical commutation relations:
\[ [a_\alpha,\mbar a_{\alpha'}]=\delta_{\alpha;\alpha'},\quad [w,\mbar w]=1,\] The
other commutators vanish: $[a_\alpha,a_{\alpha'}]=0$ and $[\mbar
w,a_\alpha]=[w,a_\alpha]=0$.

The hamiltonian of the system oscillator plus the reservoir is the sum of three
pieces: $H=H_B+\lambda H_{SB}+H_S$ where $H_B$ is the bath hamiltonian, $H_S$
the system oscillator hamiltonian and $H_{SB}$ a coupling
hamiltonian. Explicitely
\begin{eqnarray}\label{hmodel}
 H= \sum_\alpha \omega_\alpha \mbar a_\alpha a_\alpha + \lambda\big(\mbar w
\sum_\alpha\varphi_\alpha a_\alpha +w \sum_\alpha \bar \varphi_\alpha
\mbar a_\alpha\big) + \epsilon \mbar w w 
\end{eqnarray} 
We set $\hbar=1$ and $\lambda$ is a
dimensionless controlling parameter and $\varphi_\alpha$ are coupling
constants. The Hilbert space is the tensor product of the Fock spaces associated
to each of the pairs of creation-annihilation operators: ${\cal
H}=\Gamma_S\otimes\Gamma_B$ with $\Gamma_B=\bigotimes_\alpha\Gamma_\alpha$.

At initial time, the system is supposed to be prepared with a density matrix
$\rho_0\otimes\rho_B$ where the system density matrix $\rho_0$ is yet
unspecified. Density matrices are normalised, ${\rm Tr}\rho=1$. The bath density
matrix $\rho_B$ is choosen to be gaussian, $\rho_B=\bigotimes_\alpha\rho_\alpha$
with $\rho_\alpha=Z^{-1}_\alpha\, e^{-\beta_\alpha\, \mbar a_\alpha a_\alpha}$ and
$Z_\alpha^{-1}=1-e^{-\beta_\alpha}$. We shall parametrize the bath
density matrix via the mean occupation numbers $n_\alpha\equiv{\rm
Tr}_{\Gamma_\alpha}(\rho_\alpha\, \mbar a_\alpha a_\alpha)$ with
$n_\alpha=1/(e^{\beta_\alpha}-1)$.

Expectations of product of observables (which may be evaluated at different
times) are by definition given by their traces weighted by the density matrix:
$\langle{{\cal O}}\rangle\equiv \text{Tr}_{\cal H}(\rho_0\otimes\rho_B\, {\cal
O})$. As usual with composite systems, we first trace over the reservoir Hilbert
space as we are not aiming at observing it. We denote by $\mathbb{E}$ the trace
over the reservoir:
\[ \mathbb{E}[\, {\cal O}\, ] \equiv \text{Tr}_{\Gamma_B}(\rho_B\, {\cal O}\,
).\] By abuse of language we shall call $\mathbb{E}$ the expectation, it maps
general operators into operators acting on the system Hilbert space $\Gamma_S$. 
The (complete) expectations are thus given by $\langle{{\cal
O}}\rangle=\text{Tr}_{\Gamma_S}(\rho_0\, \mathbb{E}[\, {\cal O}\, ])$.

\subsection{Weak-coupling/long-time behaviour} \label{Sec:wlim} 

The aim of this section is to describe the model in the
weak-coupling/long-time/continous-density limit, corresponding to $\lambda\to0$
with $\lambda^2\, t$ fixed, in which the dynamics is known to become Markovian
\cite{Davies,Markov-limit}. The continuous-density limit ensures dissipation 
because the bath relaxation time scales as the inverse of the bath energy level spacing.
Computations presented in this Section are a bit
technical but they are needed to get an explicit description of this limiting
behaviour. The output formulas are summarised in eqs.(\ref{ztzt},\ref{2point}).
They are going to be the starting point of the construction of the quantum stochastic
process which we call quantum phase space Brownian motion. Computations may
be skipped (in a first reading) if the reader is only interested in the output.

In the Heisenberg picture, the equation of motion for an operator ${\cal O}$ is:
 \[\dot {\cal O}=i[H,{\cal O}].\]
In our model, this leads to linear equations for the oscillators:
\begin{eqnarray*}
\dot w +i\epsilon w & = & -i \lambda \sum_{\alpha} \varphi_\alpha a_\alpha\\
\dot a_\alpha +i \omega_\alpha a_\alpha & = & -i \lambda\, \bar \varphi_\alpha w 
\end{eqnarray*}
(together with their adjoints). Our aim is to understand these equations in the
weak coupling limit for an idealized bath involving a continuum of
frequencies. This can be done routinely using Laplace transform. 

Define the Laplace transform $\mtild{f}$ of a function $f$ by $\mtild{f}\equiv
\int_0^{+\infty} f(t)e^{-pt}$. Then the equations of motion become:
\begin{eqnarray*}
  p \mtild{w}(p)-w(0)+i\epsilon w(p) & = & -i \lambda \sum_{\alpha} \varphi_\alpha
  \mtild{a}_\alpha(p)\\ 
   p \mtild{a}_\alpha(p) -a_\alpha(0)+i \omega_\alpha  \mtild{a}_\alpha(p) 
   & = & -i \lambda\, \bar \varphi_\alpha \mtild{w}(p).
\end{eqnarray*}
We solve the second equation for $\mtild{a}_\alpha$ and reinject in the first to
get
\[
\left( p+i\epsilon +\lambda^2 \sum_{\alpha} \frac{|\varphi_\alpha|^2}{p+i
    \omega_\alpha}\right)\mtild{w}(p) =w(0)-i \lambda \sum_{\alpha}
\frac{\varphi_\alpha}{p+i \omega_\alpha} a_\alpha(0).
\]
As we shall discuss in details in Appendix \ref{App:weak}, the function $p+i\epsilon
+\lambda^2 \sum_{\alpha} \frac{|\varphi_\alpha|^2}{p+i \omega_\alpha}$ can only
vanish for purely imaginary values of $p$. So Laplace inversion tells that
\[
w(t)=\int_{c-i\infty}^{c-i\infty} \frac{dp}{2i\pi}\, e^{pt}\, \frac{w(0)-i \lambda
  \sum_{\alpha} \frac{\varphi_\alpha}{p+i \omega_\alpha} a_\alpha(0)}{p+i\epsilon
+\lambda^2 \sum_{\alpha} \frac{|\varphi_\alpha|^2}{p+i \omega_\alpha}}
\]
where $c$ is an arbitrary positive constant.

In the weak coupling limit the
exchanges between the bath and the system take a long time, and the short time
evolution of the system is dominated by the energy scale $\epsilon$. Once this
is factored out, the relevant times are longer. As we shall see below, if the
bath is idealized by a continuum of oscillators, the relevant long time scale
will be of order $1/\lambda^2$ and there is a non-trivial zero-coupling limit for 
\[ z(t)\equiv w( t/\lambda^2)\,e^{i\epsilon t/\lambda^2}.\]
Let us define
\[\bar{\Gamma}(p) \equiv \sum_{\alpha}
\frac{|\varphi_\alpha|^2}{\lambda^2p-i\epsilon +i\omega_\alpha}.\] After
translating and rescaling $p$ in the above formula for $w(t)$, we get a
formula for $z(t)$ which is a sum of a term proportional to the initial value
$z(0)$ and terms proportional to $a_\alpha(0)$, that is:
\[ z(t)=\bar{f}(t) z(0)+ \xi(t),\quad \text{with}\ \xi(t)\equiv \sum_{\alpha}
\bar{f}_\alpha(t) a_\alpha(0),\]
where
\begin{eqnarray*}
\bar{f}(t)&\equiv& \int_{c-i\infty}^{c-i\infty} \frac{dp}{2i\pi}\,
e^{pt}\,\frac{1}{p+\bar{\Gamma}(p)},\\
\bar{f}_\alpha(t)&\equiv& -i\lambda
 \int_{c-i\infty}^{c-i\infty} \frac{dp}{2i\pi}\,
e^{pt}\,\frac{\varphi_\alpha}{[\lambda^2 p -i\epsilon+i
  \omega_\alpha]}\frac{1}{[p+\bar{\Gamma}(p)]}
 \end{eqnarray*}
Taking complex conjugates, we note that 
\[ f_\alpha(s)=i\lambda \int_{c-i\infty}^{c-i\infty} \frac{dq}{2i\pi}\,e^{qs}\,
\frac{\bar{\varphi}_\alpha}{[\lambda^2 q +i\epsilon-i
  \omega_\alpha]}\frac{1}{[q+\Gamma(\bar q)]}.\]

Because the bath is Gaussian and $\xi(t)$ is linear in bath oscillators, its
properties are fully specified by the value of the commutator
$[\xi(t),\mbar{\xi}(s)]$ and the two point function $\expect[
\mbar{\xi}(s)\xi(t)]$. From the basic oscillator commutation relations, we have
$[\xi(t),\mbar{\xi}(s)] = \sum_{\alpha} \bar{f}_\alpha(t)f_\alpha(s)$. Using the
identity
$\frac{1}{p+\omega}\frac{1}{q-\omega}=\frac{1}{p+q}\left(\frac{1}{p+\omega}+
  \frac{1}{q-\omega}\right)$ we get
\begin{eqnarray*} [\xi(t),\mbar{\xi}(s)] 
=\int_{c-i\infty}^{c-i\infty} \frac{dp}{2i\pi}\frac{dq}{2i\pi}\, e^{pt} e^{qs}\, 
\frac{\bar{\Gamma}(p)+\Gamma(\bar q)}{[p+\bar{\Gamma}(p)][q+\Gamma(\bar q)][p+q]}.
\end{eqnarray*}
In the same vein, we find $\expect [\mbar{\xi}(s)\xi(t)] = \sum_{\alpha}
n_{\alpha}\bar{f}_\alpha(t)f_\alpha(s)$, which can be written as:
\begin{eqnarray*} 
\expect [\mbar{\xi}(s)\xi(t)] 
=\int_{c-i\infty}^{c-i\infty}\frac{dp}{2i\pi}\frac{dq}{2i\pi}\,e^{pt} e^{qs}\,
\frac{\bar{\Upsilon}(p)+ \Upsilon(\bar q)}{[p+\bar{\Gamma}(p)][q+\Gamma(\bar
  q)[]p+q]}, 
\end{eqnarray*}
where 
\[\bar{\Upsilon}(p)\equiv \sum_{\alpha}
\frac{n_{\alpha}|\varphi_\alpha|^2}{\lambda^2p-i\epsilon +i\omega_\alpha}.\]

Turning now to the situation when there is a density of oscillators, we
replace $\sum_{\alpha} |\varphi_\alpha|^2 \delta (\omega-\omega_\alpha)$ by
$r(\omega)$ and, interpreting  $n_\alpha$ as $n(\omega_\alpha)$, we find:
\[ \bar{\Gamma}(p)=\int_0^{+\infty} d\omega
\frac{r(\omega)}{\lambda^2p-i\epsilon +i\omega},\;
\bar{\Upsilon}(p)=\int_0^{+\infty} d\omega
\frac{n(\omega)r(\omega)}{\lambda^2p-i\epsilon +i\omega}.\] 
Taking naively the limit $\lambda^2 \rightarrow 0^+$ and remembering that the
integration contours stay in the half-plane $\Re e \, p >0$, we find that
$\bar{\Gamma}(p) \rightarrow \bar{\gamma}$ with
\begin{equation}\label{defga}
 \bar{\gamma}\equiv \bar{\Gamma}(0^+)=\pi
r(\epsilon)+i\fint_0^{+\infty} d\omega \frac{r(\omega)}{\epsilon -\omega}.
\end{equation}
Analogously, 
$\bar{\Upsilon}(p)+\Upsilon(\bar q) \rightarrow 2\pi n(\epsilon) r(\epsilon)$.

Thus, in the weak coupling limit we get:
\begin{eqnarray*} 
\bar{f}(t) & = & \int_{c-i\infty}^{c-i\infty} \frac{dp}{2i\pi}\, e^{pt}\,
   \frac{1}{p+\bar{\gamma}},\\
  \left[\xi(t),\mbar{\xi}(s)\right] & = & 2\pi 
  r(\epsilon)\int_{c-i\infty}^{c-i\infty} \frac{dp}{2i\pi}\frac{dq}{2i\pi}\,e^{pt}
  e^{qs}\,\frac{1}{[p+\bar{\gamma}][q+\gamma][p+q]}, \\
   \expect[ \mbar{\xi}(s)\xi(t)] & = & 2\pi n(\epsilon) r(\epsilon)
   \int_{c-i\infty}^{c-i\infty} \frac{dp}{2i\pi}\frac{dq}{2i\pi}\,e^{pt}
   e^{qs}\,\frac{1}{[p+\bar{\gamma}][q+\gamma][p+q]} .
   \end{eqnarray*}
   Note that $\expect[ \mbar{\xi}(s)\xi(t)]
   =n(\epsilon)\left[\xi(t),\mbar{\xi}(s)\right]$.

The remaining integrals are obtained by application of the residue theorem.
For instance, $\bar{f}(t)=e^{-\bar{\gamma}t}$.
The commutator requires a slight discussion. If $t \geq s \geq 0$,
\[\int_{c-i\infty}^{c-i\infty} \frac{dq}{2i\pi} e^{qs}\frac{1}{q+\gamma}
\frac{1}{p+q}= \frac{e^{-ps}-e^{-\gamma s}}{\gamma-p},\]
because the integration contour can be pushed to the
left. This has no pole at $p=\gamma$. Then,
\[\int_{c-i\infty}^{c-i\infty} \frac{dp}{2i\pi}e^{pt}\frac{1}{p+\bar{\gamma}}
\,\frac{e^{-ps}-e^{-\gamma s}}{\gamma-p}=\frac{e^{-\bar{\gamma}(t-s)}-e^{-\gamma s
    -\bar{\gamma} t}}{\gamma + \bar{\gamma}},\]
because the integration contour can be pushed to the left again. The case  $s
\geq t \geq 0$ is treated analogously. 

Hence, defining 
\begin{eqnarray*}
G(t,s)& \equiv & e^{i(t-s)\Im m \, \gamma}\left(e^{-|t-s| \Re 
    e\,\gamma}-e^{-(t+s) \Re e\,\gamma}\right),
\end{eqnarray*}
we end up with $f(t)=e^{-\bar{\gamma}t}$ and
\begin{eqnarray*}
  z(t) & = & e^{-\bar{\gamma}t} \, z(0)+ \xi(t),\\ \left[ \xi
    (t),\mbar{\xi}(s)\right] & = & 
  G(t,s) \\ \expect[ \mbar{\xi}(s)\xi(t)] & = & n(\epsilon) G(t,s).
\end{eqnarray*}

The above derivation of the long-time/weak-coupling/continuous-density limit
relies heavily on the magic of contour deformation. Though the final result,
damping when the bath is infinite, is physically satisfactory, it may seem
disturbing that taking the limit of a function with many poles but all purely
imaginary, one finds functions with single poles having a negative real part.
The model we consider is simple enough that one can make a more down-to-earth
derivation which shows more clearly that damping arises from destructive
interferences. It also leads to an explicit spectral representation (\ref{ztmich}) 
of the process directly inherited from the discrete harmonic bath oscillator decomposition. 
 This is presented in Appendix \ref{App:weak}.

\section{Quantum phase space Brownian motion} \label{Sec:brown}
Let us reformulate (and summarise)
the output of the weak-coupling/long-time limit of previous section. The time
evolutions of the system canonical operators $z$ and $\mbar z$ are:
\begin{eqnarray}\label{ztzt} z(t)= e^{-\bar \gamma t}z+\xi(t)\ ,\quad \mbar
z(t)= e^{-\gamma t}\mbar z+\mbar\xi(t)
\end{eqnarray} with $\gamma$ a complex parameter,
$\Re\text{e}\,\gamma\geq0$. We shall set $\gamma=\kappa+i\nu$ 
and $\bar\gamma=\kappa-i\nu$. The
fields $\xi(t)$ and $\mbar\xi(t)$, which represent "the quantum noice", are
linear combinations of the reservoir  creation-annihilation operators.
Commutation relations are: 
\[ [z,\mbar z]=1\] and
\begin{eqnarray}\label{2point}
 [\xi(t),\mbar \xi(s)]= G(t,s)\equiv
	\begin{cases} (e^{\gamma t}-e^{-\bar \gamma t})e^{-\gamma s}\ ; &
\text{for}\ t\leq s \\ e^{-\bar \gamma t}(e^{\bar \gamma s}-e^{-\gamma
s})\ ; & \text{for}\ t\geq s \end{cases} 
\end{eqnarray}
The system operators commute with
the reservoir operators so that $[z,\xi(t)]=[\mbar z,\xi(t)]=0$.

The measure $\mathbb{E}$, induced by tracing over the reservoir degrees of
freedom with the bath density matrix, is such that $\mbar\xi(t)$ and $\xi(s)$
are gaussian with two-point function:
\[ \mathbb{E}[ \mbar \xi(t)\,\xi(s)]= \mathfrak{n}_0\, G(s,t) \] with
$\mathfrak{n}_0$ real positive.  In previous Section \ref{Sec:wlim}, this
parameter was denoted $n(\epsilon)$. Since physical phenomena concentrate at the
energy scale $\epsilon$ in the long-time/weak-coupling limit, all details of the
reservoir parameters have been erased and summarised in the very few parameters
$\gamma$, $\mathfrak{n}_0$ and in the form of the commutation relations $G(t,s)$
and in the two-point functions $\mathfrak{n}_0\, G(s,t)$.

The aim of this section is to extract (and illustrate) the structure of quantum
processes in this particular example which possesses all generic properties plus
some peculiar ones. It has most properties for being a quantum analogue of a
(two-dimensional) Brownian motion. In particular its increments are independent
and identically distributed.

\subsection{Flow and expectations} 
$\bullet$ {\it Algebras:} We first have to
identify where the flow takes place.  Let $A_0$ be the algebra generated by the
system canonical operators $z$ and $\mbar z$, and let $B_{[0,t]}$ be the
algebra generated by all the $\xi(s)$ and $\mbar \xi(s)$ for $s\in[0,t]$. We
set
\begin{eqnarray}\label{algebre} A_t\equiv A_0\otimes B_{[0,t]}\ ,\quad
A_\infty\equiv A_0\otimes B_{[0,\infty)}.
 \end{eqnarray}

 Although understandable by ``common sense'' and useful in practice, the
 definition of $B_{[0,t]}$ as ``the algebra generated by the $\xi(s)$'s with
 $s\leq t$'' is not precise. Thus, let $\mathbb{L}^2_\kappa(\mathbb{R}_+)$ be
 the Hilbert space of functions $s\to f(s)$, such that $e^{\kappa s}f(s)$ is
 square integrable on the positive real line, equipped with the scalar product
 $(g|f)_\kappa=\int_0^\infty ds\, \overline{g(s)}\, 2\kappa e^{2\kappa s}\,
 f(s)$. Let $\Gamma(\mathbb{L}^2_\kappa(\mathbb{R}_+))$ be the Fock space over
 $\mathbb{L}^2_\kappa(\mathbb{R}_+)$, and let $\mathfrak{V}_f$ be the canonical operators
 on $\Gamma(\mathbb{L}^2_\kappa(\mathbb{R}_+))$ depending linearly on functions
 $f$ with commutation relations $[\mathfrak{V}_f,\mbar{\mathfrak{V}}_g]=(g|f)_\kappa$. Then
 \[ \xi(t) = e^{-\bar\gamma t}\,\mathfrak{V}_{{\bf 1}_{[0,t]}}.\] More explicitely, the
 bath operators $\xi(t)$ may be represented as (recall that
 $\gamma=\kappa+i\nu$):
 \[ \xi(t) = e^{-\bar \gamma t} \int_0^t du\, (2\kappa e^{2\kappa
   u})^{\frac{1}{2}}\, a(u),\] where $a(s)$ and $\mbar a(s)$ are bare canonical
 operators with commutation relations:
\[ [a(s),a(s')]=0,\quad [a(s),\mbar a(s')]=\delta(s-s').\] 
One may use this
representation of $\xi(t)$ to write a "quantum stochastic differential equation"
for $z(t)$.  Indeed differentiating eq.(\ref{ztzt}), we get:
\begin{eqnarray}\label{qdiif} 
  dz(t)=-\bar\gamma\, z(t)\, dt + \sqrt{2\kappa}\,e^{i\nu t}\, d\mathfrak{A}(t),
 \end{eqnarray} 
 where we set $d\mathfrak{A}(t)\equiv \int_t^{t+dt}a(u)du$.  
 A similar representation is of course at the basis of the quantum stochastic 
 calculus due to Hudson and Parthasarathy \cite{Huds-parth}. 

 Let $B_\infty\equiv B(\Gamma(\mathbb{L}^2_\kappa(\mathbb{R}_+))$ be the algebra
 of bounded operators on $\Gamma( \mathbb{L}^2_\kappa(\mathbb{R}_+)$, which we
 identify with $B_{[0,\infty)}$. Recall that $\Gamma( \mathbb{L}^2_\kappa(\mathbb{R}_+)=\Gamma(\mathbb{L}^2_\kappa([0,t])\otimes \Gamma(\mathbb{L}^2_\kappa(]t,\infty))$.
 We may then define $B_{[0,t]}$ as the $B_\infty$-subalgebra (this is more an embedding than an inclusion):
 \[ B_{[0,t]}\equiv B\big( \Gamma(\mathbb{L}^2_\kappa([0,t])) \subset
 B_\infty.\] Notice that $B_{[0,s]}\subset B_{[0,t]}$ for $s<t$ and
 $B_\infty=B_{[0,t]}\otimes B_{]t,\infty)}$.  There is some freedom in the
 choice of the algebras $B_{[0,t]}$, the above choice is a minimal one, maybe
 not adapted to all situations one may encounter. However, most of the following
 computations are algebraic and valid for a larger class of operators on
 $\Gamma( \mathbb{L}^2_\kappa(\mathbb{R}_+))$ than those in $B_{[0,t]}$.
 \medskip

 \noindent $\bullet$ {\it Flow:} The time evolution of the system operator
 defines a flow $J_t$ from $A_0$ to $A_t$ by:
\begin{eqnarray}\label{flow} J_t:A_0\to A_t,\quad J_t(z)\equiv z(t)=e^{-\bar 
    \gamma t}z+\xi(t).
\end{eqnarray} Similarly, $J_t(\mbar z)\equiv \mbar z(t)=e^{-\gamma
  t}\mbar z+\mbar \xi(t)$.  It is a $*$-homomorphism since $[z(t),\mbar
z(t)]=1$, because $G(t,t)=1-e^{-(\gamma+\bar\gamma)t}$, so that
\[ J_t(e^{\mu \mbar z} e^{\bar\mu z})=e^{\mu \mbar z(t)} e^{\bar\mu
z(t)} =e^{\mu(t) \mbar z} e^{\bar\mu(t) z}\,e^{\mu\mbar\xi(t)}
e^{\bar\mu\xi(t)}\, . \] with \[\mu(t)=\mu\, e^{-\gamma t},\quad
\bar\mu(t)=\bar\mu\, e^{-\bar\gamma t}.\] Eq.(\ref{flow}) is the quantum
analogue of a stochastic classical flow with $\xi(t)$ playing the role of
quantum noise.  
\medskip

\noindent $\bullet$ {\it Measure:} The quantum noises $\xi(t)$ are elements of
$B_\infty$. The measure over it is $\mathbb{E}$. It is gaussian with two-point
function $\mathbb{E}[\mbar\xi(t)\xi(s)]=\mathfrak{n}_0G(s,t)$ so that
\begin{eqnarray*} \mathbb{E}[\,
e^{\mu_1\mbar\xi(t_1)}e^{\bar\mu_1\xi(t_1)}\cdots
e^{\mu_N\mbar\xi(t_N)}e^{\bar\mu_N\xi(t_N)}\, ] = \exp[\sum_{i,j}
\bar\mu_i\mu_j G(t_i,t_j)(\mathfrak{n}_0+{\bf 1}_{\{i<j\}})] .
\end{eqnarray*} 
Here $\mu_j$ and $\bar\mu_j$ are formal parameters not necessary complex
conjugate. If one chooses $\bar \mu_j=-\mu_j^*$, then
the operators involved in the above equation are in $B_\infty$.

The measure $\mathbb{E}$ is neutral with respect to elements of the system
oscillator algebra, that is $\mathbb{E}[a_0\, b]=a_0 \mathbb{E}[b]$ for any
$a_0\in A_0$, so that it extends to a map from $A_\infty$ to $A_0$:
\[ \mathbb{E}: A_\infty\to A_0. \] On the bare canonical operators $a(s)$ the
measure is $\mathbb{E}[\mbar a(s)a(s')]=\mathfrak{n}_0\delta(s-s')$, and:
\[ \mathbb{E}[d\mathfrak{A}(t)\,d\mathfrak{A}(t)]=0,\quad
\mathbb{E}[d\mbar{\mathfrak{A}}(t)\,d\mathfrak{A}(t)]=\mathfrak{n}_0\, dt.\]

To complete the setting, one has to provide the measure
$\text{Tr}_{\Gamma_S}(\rho_0\cdots)$ on $A_0$ specified by the system density
matrix. Classically, $\mathbb{E}$ would be the measure on the noise and
$\text{Tr}_{\Gamma_S}(\rho_0\cdots)$ the measure on the initial position of the
stochastic process.

\subsection{Dynamical maps and density matrices} 
$\bullet$ {\it Dynamical map:} The map
$\Phi_t$ on $A_0$ is obtained by evaluating one-point functions:
$\Phi_t(a)=\mathbb{E}[J_t(a)]$ for $a\in A_0$. In the present case, it yields:
\begin{eqnarray}\label{stomap} \Phi_t(e^{\mu \mbar z} e^{\bar\mu z})=
e^{\mu(t) \mbar z} e^{\bar\mu(t) z}\,
e^{\mathfrak{n}_0[\mu\bar\mu-\mu(t)\bar\mu(t)]}.
 \end{eqnarray} 
 Of course $\Phi_t(1)=1$. As can be checked by a direct computation, it
 defines a semi-group on $A_0$: $$\Phi_t\circ\Phi_s=\Phi_{t+s}.$$ The form of
 the generator of the dynamical maps that we are going to find, in
 eq.(\ref{lind}) below, implies that $\Phi_t$ are completely positive maps, as
 required by the theory of quantum processes \cite{Barch,Biane1,Attal}.  It is the
 quantum analogue of transition probability kernels for classical stochastic
 processes.  \medskip

 \noindent $\bullet$ {\it Lindbladian:} The Lindbladian is the generator of
 the dynamical semi-group.  Indeed, the semi-group law (with extra continuous
 properties) implies that
\[\Phi_t=\exp(\,tL\, )\] 
where $L=\frac{d}{dt}\Phi_t\vert_{t=0}$ is a so-called Lindbladian,
with a structure imposed by the general theory of completely positive
dynamical maps \cite{Lindblad}:
\[ L(a) = i[h_s,a] + \sum_j g_j\, (2D_ja\mbar D_j- D_j\mbar D_j a - a D_j \mbar
D_j),\quad a\in A_0\] 
with some effective hamiltonian $h_s=\mbar h_s$ and some operators $D_j$
and $g_j>0$.  In the present case, $L$ is quadratic in $z$ and $\mbar z$ and,
for all $a\in A_0$,
\begin{eqnarray} \label{lind}
 L(a) =- i\nu\, [\mbar z z,a] + g_1\, (2za\mbar z- z\mbar z a- a z\mbar
z) + g_2\, (2\mbar z a z - \mbar z z a - a \mbar z z),
\end{eqnarray}
where we set $\gamma=\kappa+i\nu$ and
\[ g_1 = \kappa\mathfrak{n}_0,\quad g_2= \kappa(\mathfrak{n}_0+1).\] Of course
$L(1)=0$ and $L(\mbar a)=\mbar {L(a)}$.  Set $L(a)=-i\nu[\mbar z z,a]+D(a)$ where
$D(a)$ is the so-called dissipative part of the Linbladian. It may alternatively be
written as
\[ D(a)= -\kappa\mathfrak{n}_0\,\big( [\mbar z,[z,a]]+[z,[\mbar
z,a]]\big)+\kappa\, (2\mbar z a z - \mbar z z a - a \mbar z z) .\] The double
commutator term coincides with the (quantum analogue of the) two dimensional
Laplacian.  As expected, the imaginary part of $\gamma$ only enters into the
hamiltonian part of the Linbladian, but note that the frequency $\nu$ is not the
bare frequency $\epsilon$. The dissipative part of $L$ only depends on $\re
\gamma$ and $\mathfrak{n}_0$. 
This Lindbladian has been considered in ref.\cite{Breuer-Pet} 
to describe damped harmonic oscillator.
\medskip

\noindent $\bullet$ {\it Master equation:} This refers to the evolution equation
for the system reduced density matrix. It is obtained from the observable time
evolution by duality since ${\rm Tr}_{\Gamma_S}( \rho_0\, \Phi_t(a) )= {\rm
Tr}_{\Gamma_S}(\rho_t\, a)$ for $a\in A_0$. Hence
$\rho_t=e^{tL^*}\cdot\rho_0$. So that
\begin{eqnarray}\label{dualL} \frac{d}{dt}\rho_t = L^*\cdot\rho_t
\end{eqnarray} with
\[ L^*\cdot\rho= i\nu[\mbar zz,\rho]+ g_2\, (2z\rho\mbar z- \rho\mbar z z- \mbar
z z \rho) + g_1\, (2\mbar z \rho z - \rho z\mbar z - z\mbar z \rho) .\]
Eq.(\ref{dualL}) is the quantum analogue of Fokker-Planck equations for Markov
processes. It preserves the normalisation of $\rho$, i.e. $\text{Tr}_{\Gamma_S}(
L^*\cdot\rho )=0$.

The system admits an invariant measure which is a density matrix
$\rho_\text{inv}$ such that $L^*\cdot\rho_\text{inv}=0$. It is gaussian:
\[ \rho_\text{inv}= \frac{1}{Z}\, e^{-\sigma\, \mbar z z},\quad \text{with}\
e^\sigma= \frac{\mathfrak{n}_0+1}{\mathfrak{n}_0},\] and $Z=1/(1-e^{-\sigma})$.
Alternatively, $\mathfrak{n}_0=1/(e^{\sigma}-1)$. An effective temperature may be defined either
by $\sigma=\nu/T$ or $\sigma=\epsilon/T$.  The invariant measure depends on
$\mathfrak{n}_0$ but not on $\kappa$.

Since $\kappa$ represents damping effects, the approach to equilibrium is
$\kappa$-dependent. It is exponentially fast as $e^{-2\kappa\, t}$ with
relaxation time $1/\kappa$. It is governed by the smallest eigenvalue of $L^*$
which is $-2\kappa$:
\[ L^*\cdot\rho_1= -2\kappa\, \rho_1,\] with eigenvector $\rho_1=e^{-\sigma\,
  \mbar z z}[\mbar z z -2\mathfrak{n}_0]$. Other eigenvalues of $L^*$ (in the
zero charge sector) are $-2p\kappa$ with $p$ integer and eigenvector
$\rho_p=e^{-\sigma\, \mbar z z}P_p(\mbar z z)$ with $P_p$ polynomial of degree
$p$.

The invariant measure is stable since all eigenvalues of $L^*$ are non positive.
This is a consequence of the complete positivity of the dynamical maps which
implies that the dissipative part of $L^*\cdot\rho$ is the sum of terms of the
form $g_j(2D_j\rho\mbar D_j-\rho\mbar D_j D_j - \mbar D_j D_j \rho)$ which are
all non positive operators for $g_j>0$. Indeed $\text{Tr}_{\Gamma_S}(\rho\,
(L^*\cdot\rho))$ is the sum of $2g_j(\text{Tr}_{\Gamma_S}(\rho D_j\rho \mbar
D_j)- \text{Tr}_{\Gamma_S}(\rho D_j\mbar D_j \rho))\leq 0$ by the Cauchy-Schwarz
inequality.

\subsection{Quantum filtration and conditional expectations} $\bullet$ {\it
Filtration:} The set of embedded algebras $A_t=A_0\otimes B_{[0,t]}$ defines a
filtration~\footnote{We do not enter here in the subtleties associated to
left/right time limits as the construction is clearly time continuous.}:
\[ A_0\subset A_s \subset A_t \subset A_\infty,\quad \text{for}\ 0<s<t<\infty.\]
The algebra $A_s$ has to be understood as the past of the process up to time
$s$. The image $J_s(A_0)$ of the system algebra by the flow is identified as the
present of the process at time $s$: $J_s(A_0)\subset A_s$. With these
definitions the past includes the present. The algebra $A_0\otimes
B_{]t,\infty)}$ may be identified with the futur of the process.  
\medskip

\noindent $\bullet$ {\it Conditional expectations:} These are projectors
$\mathbb{E}_s$ from $A_\infty$ into $A_s$ such that
\begin{eqnarray}\label{EsEs} \mathbb{E}_s:\, A_\infty\to A_s,\quad
\mathbb{E}_{s_1}\circ \mathbb{E}_{s_2} = \mathbb{E}_{\text{min}(s_1,s_2)}, \quad
\text{for}\ s_1,s_2\geq 0,
\end{eqnarray} and compatible with the flow and the stochastic map in the sense
that, for all $a\in A_0$:
\begin{eqnarray}\label{EsJt} 
\mathbb{E}_{s}[\, J_t(a)\, ] = J_s[\,
\Phi_{t-s}(a)\, ]\ ,\quad \text{for}\ t>s\geq 0.
\end{eqnarray} 
$\mathbb{E}_s$ has to be understood as the expectation
conditioned on the past of the process up to time $s$. 
Eq.(\ref{EsJt}) expresses the Markov property of the process.
In the present case, compatibility with the flow imposes that, for $t>s$,
\[ \mathbb{E}_s[\,e^{\mu \mbar\xi(t)}e^{\bar\mu\xi(t)}\,] = e^{\mu(t-s)
\mbar\xi(s)}e^{\bar\mu(t-s) \xi(s)}\,
e^{\mathfrak{n}_0[\mu\bar\mu-\mu(t-s)\bar\mu(t-s)]}.\] 
This defines
$\mathbb{E}_s$ on single-time operator but we have to define it on product of
multi-time operators. Since we know the commutation relations in $A_\infty$ it
is enough to define it on product of time ordered operators. That is, we have to define
\[ \mathbb{E}_s[\, e^{\mu_1 \mbar\xi(t_1)}e^{\bar\mu_1\xi(t_1)}\cdots e^{\mu_N
  \mbar\xi(t_N)}e^{\bar\mu_N\xi(t_N)}\,] \] 
  with $t_1<\cdots<t_N$. A
constructive way to do it consists in (recursively) imposing that $\mathbb{E}_s$
is neutral with respect to left multiplication by elements of $A_s$, i.e.
$\mathbb{E}_s[ab]=a\mathbb{E}_s[b]$ for $a\in A_s$.  After a few computations
summarised in Appendix \ref{App:proof}, we get:
\begin{eqnarray}\label{defEs} && ~~~~~~~ \mathbb{E}_s[ e^{\mu_1
\mbar\xi(t_1)}e^{\bar\mu_1\xi(t_1)}\cdots e^{\mu_N
\mbar\xi(t_N)}e^{\bar\mu_N\xi(t_N)}\,]\\ &=&
e^{\mu_1(t_{1;s})\mbar\xi(s)}e^{\bar\mu_1(t_{1;s})\xi(s)}\cdots
e^{\mu_N(t_{N;s}) \mbar\xi(s)}e^{\bar\mu_N(t_{N;s})\xi(s)}\,
e^{X^{(N)}_s}\, e^{\mathfrak{n}_0Y^{(N)}_s} \nonumber
\end{eqnarray} for $s<t_1<\cdots<t_N$, with
\begin{eqnarray*} X^{(N)}_s&=& \sum_{i<j}
\bar\mu_i\mu_j(t_{j;i})-\sum_{i<j}\bar\mu_i(t_{i;s})\mu_j(t_{j;s}) \\
Y^{(N)}_s&=& \sum_i \bar\mu_i\mu_i+ \sum_{i<j}
[\bar\mu_i\mu_j(t_{j;i})+\mu_i\bar\mu_j(t_{j;i})]-\sum_{i,j}\bar\mu_i(t_{i;s})\mu_j(t_{j;s})
\end{eqnarray*} 
where $t_{j;i}=t_j-t_i$ and $t_{j;s}=t_j-s$. Notice that
the r.h.s. of eq.(\ref{defEs}) only involves $\xi(s)$ and $\mbar\xi(s)$. One checks that
$\mathbb{E}_{s=0}=\mathbb{E}$ as it should be.
\medskip

\noindent $\bullet$ {\it Markov and other properties:}. The previous definition
is designed to ensure that relations (\ref{EsEs}) and (\ref{EsJt}) hold, but this can
be checked directly. See Appendix \ref{App:proof}.

Since it only involves $\xi(s)$ and $\mbar\xi(s)$, it also satisfies a (weak)
Markov property because the image by $\mathbb{E}_s$ of the future algebra
$B_{]s,\infty)}$ maps into the present algebra $J_s(A_0)$:
\begin{eqnarray}\label{markov} \mathbb{E}_s: B_{]s,\infty)} \to J_s(A_0)
\end{eqnarray}

By construction, $\mathbb{E}_s$ is neutral with respect to left multiplication
by elements in $A_s$, but remarkably it is also neutral with respect to right
multiplication. Namely,
\begin{eqnarray}\label{neutral} \mathbb{E}_s[ab]=a\,\mathbb{E}_s[b]\ ,\quad
\mathbb{E}_s[ba]=\mathbb{E}_s[b]\,a\ , \quad \text{for}\ a\in A_s
\end{eqnarray} for all $b\in A_\infty$. The first relation is true by
construction and the second one follows from the relation
\[ \mu\, [\xi(t_0),\mbar\xi(t)]=\mu(t-s)\, [\xi(t_0),\mbar\xi(s)],\] 
for $t_0\leq s\leq t$. See Appendix \ref{App:proof}.
This property is remarkable. In particular it means that we
would have got the same conditional expectations if we would have constructed it
using neutrality under right multiplication instead of neutrality under left
multiplication as we did. In this sense, the conditional expectation is unique.

By construction (also), the conditional expectation satisfies the nested
formula, similar to that valid in the classical theory, but for time ordered
products $s<t_1<\cdots<t_N$:
\begin{eqnarray}\label{nested} 
  && ~~~~~ \mathbb{E}_s[\,J_{t_1}(a_1)\cdots J_{t_{N-1}}(a_{N-1}) J_{t_N}(a_N)\,
  ]\\  
  &=& J_s\big[ \Phi_{t_1-s}(a_1\Phi_{t_2-t_1}(\cdots a_{N-1}
  \Phi_{t_N-t_{N-1}}(a_N))\big]  \nonumber
\end{eqnarray}
for $a_j\in A_0$. Note that the nesting is ordered from right to left. A similar
nested formula applies for an opposite ordering of the time but with an opposite
nesting. Both formula follow from eqs.(\ref{EsEs}) and from the left and right
neutralities.

The nested formula eq.(\ref{nested}) is of course an echo that the bath dynamics
has been frozen by the infinite-volume/continuous-density limit.  \medskip

\noindent $\bullet$ {\it Martingales:} As for classical processes, we may define
martingales -- which, for instance, are instrumental for computing
probabilities of events related to stopping times. A family of elements $M_t$
in $A_t$ is called martingale if
\[ \mathbb{E}_s[\, M_t\,]= M_s,\quad \text{for}\ s<t.\]
Examples are for instance given by:
\[ M_t= e^{\mu e^{\gamma t}\mbar\xi(t)}\, e^{\bar\mu e^{\bar\gamma t}\xi(t)}\,
e^{-\mathfrak{n}_0\mu\bar\mu e^{(\gamma+\bar\gamma)t}}.\] To first order in
$\bar\mu$, this gives $e^{\bar\gamma t}\xi(t)$ as a martingale.

\subsection{Time evolution in $A_\infty$} $\bullet$ {\it Flow:} By
eq.(\ref{flow}) we have defined the flow of system observables. We may extend
this flow to elements of $A_\infty$. That is, we may define a semi-group of
$A_\infty$-endomorphisms, $\sigma_s$ with $\sigma_0=\text{Id}$, such that:
\begin{eqnarray}\label{Aflow} \sigma_s:\ A_\infty\to A_\infty,\quad
\sigma_s\circ\sigma_t=\sigma_{s+t},\quad \sigma_s\circ J_t=J_{t+s},
 \end{eqnarray} for $ s,t\geq 0$. The second relation, namely $\sigma_s\circ
J_t=J_{t+s}$, means that $\sigma_s$ extends the flow $J_s$. In particular
$\sigma_s$ acts like $J_s$ on elements of $A_0$. It implies that:
\begin{eqnarray}\label{sigma} \sigma_s(z)\equiv e^{-\bar\gamma
s}z+\xi(s),\quad \sigma_s(\xi(t)¤)\equiv \xi(t+s)-e^{-\bar\gamma t}\xi(s).
 \end{eqnarray} Similarly $\sigma_s(\mbar z)=J_s(\mbar z)$ and
$\sigma_s(\mbar\xi(t))=\mbar\xi(t+s)-e^{-\gamma t}\mbar\xi(s)$. One may
check by a direct computation that $\sigma_s$ are $*$-endomorphisms and
\[
[\,\sigma_s(\xi(t)),\sigma_s(\mbar\xi(t'))\,]=[\,\xi(t),\mbar\xi(t')\,]=G(t,t'),\]
for all $t,\, t'\geq0$. Of course, $\sigma_s\circ\sigma_t=\sigma_{s+t}$. Since
both $\sigma_s$ and $J_t$ are morphisms, the compatibility relation
$\sigma_s\circ J_t=J_{t+s}$ is valid on $A_0$ because it is true for the
generators of $A_0$.  \medskip

\noindent $\bullet$ {\it Covariance:} The flow $\sigma_t$ is compatible with the
conditional expectations in the sense that
\begin{eqnarray}\label{cov}
\sigma_t\circ\mathbb{E}_{s}=\mathbb{E}_{t+s}\circ\sigma_t .
\end{eqnarray} This is a direct consequence of the previous
construction. Indeed, neutrality of $\mathbb{E}_s$ for elements of $A_s$ implies
that this is true on $A_\infty$ if it is true when applied to elements of the
future algebra $B_{]s,\infty)}$. To check it on $B_{]s,\infty)}$, it is enough
to check that it holds true for product elements of the form $J_{t_1}(a_1)\cdots
J_{t_N}(a_N)$ with $a_j\in A_0$ and times ordered $s<t_1<\cdots<t_N$. On such
elements, eq.(\ref{cov}) is a direct consequence of the nested formula
eq.(\ref{nested}) for conditional expectations and of the compatibility relation
$\sigma_t(J_s(a))=J_{t+s}(a)$ on $A_0$.

\subsection{Independence of increments} We call
$\sigma_s(\xi(t))=\xi(t+s)-e^{-\bar\gamma t}\xi(s)$ the increment from $s$ to
$t+s$.

Since $\sigma_s$ is an endomorphism and since the two-point functions are
proportional to the commutators $G(t,t')$, the increments are identically
distributed:
\begin{eqnarray*} 
\mathbb{E}[\, \sigma_s(\mbar\xi(t))\,\sigma_s(\xi(t'))\,]&=&\mathbb{E}[\,
\mbar\xi(t)\,\xi(t')\,]=\mathfrak{n}_0\,G(t',t)
\end{eqnarray*} for all $s>0$.

Furthermore, the relation
$[\xi(t_0),\mbar\xi(t+s)]=\mu(t)\,[\xi(t_0),\mbar\xi(s)]$ for $t_0\leq s$
and $t\geq0$, which was instrumental in proving the left/right neutrality of the
conditional expectations, implies that:
\begin{eqnarray}\label{iid} [\, \sigma_s(\xi(t))\,,\,\mbar\xi(t_0)\,]
&=&0,\quad \text{for}\ t_0\leq s \\ \mathbb{E}[\,\sigma_s(\xi(t))\,\mbar
\xi(t_0)\,] &=&0,\quad \text{for}\ t_0\leq s \nonumber
\end{eqnarray} These two relations then imply that $[\, \sigma_s(\xi(t)),a]
=0$ and $\mathbb{E}[\,\sigma_s(\xi(t))\,a\,] =0$ for any $a\in A_s$. It means
that the increments commute with the past algebra $A_s$ and are independent of
the past. Another remarkable property.

\section{Limiting cases  and decoherence} \label{Sec:class}
\subsection{The quantum $t$-Brownian limit}  
The process simplifies in the limit $\gamma\to 0$. Set $\gamma=\kappa+i\nu$ and
let $\chi(t)=\frac{1}{\sqrt{\kappa}} \xi(t)e^{-i\nu t}$. In the limit
$\kappa\to0$, these can be written as $\chi(t)= \int_0^tds\, a(s),$ in terms of
the bare canonical operators $a(s)$, with $[a(s),\mbar a(s')]=\delta(s-s')$.
Then
\begin{eqnarray*}
[\, \chi(t),\mbar\chi(s)\,]&=&2\, \text{min}(t,s)\\
\mathbb{E}[\mbar\chi(t)\chi(s)]&=&2\mathfrak{n}_0\,\text{min}(t,s)
\end{eqnarray*}
in the limit $\kappa\to 0$. The real and imaginary parts of $\chi(t)$ are two
Brownian motions but which do not commute except at time $t=0$.

This coincides with the quantum process (also called a quantum Brownian motion)
defined in ref.\cite{Biane} using Hopf algebra techniques. Actually, our initial
algebra $A_0$ has to be slightly modified in order to take care of the limit
$\kappa\to0$. Following ref.\cite{Biane}, we define the initial algebra as the
Heisenberg algebra $\tilde A_0$ generated by elements $\tilde z,\ \mbar{\tilde
  z}$ and $\tau$ with commutation relations
\[ [\tilde z, \mbar{\tilde z}]=\tau,\quad [\tau, \mbar{\tilde z}]=0=[\tau,
{\tilde z}].\] The flow is defined by $J_t(\tilde z)=\chi(t)$ and $J_t(\tau)=t$.
It is a $*$-homomorphism. The dynamical maps is the $\gamma\to0$ limit of
eq.(\ref{stomap}), namely:
\[\Phi_t(e^{\eta \mbar{\tilde z}} e^{\bar\eta \tilde z})=
e^{\eta \mbar{\tilde z}} e^{\bar\eta \tilde z}\,e^{2\eta\bar\eta\,\mathfrak{n}_0 t}.\] 
  In this limit, the conditional expectations are
\[ \mathbb{E}_s[ e^{\eta\mbar\chi(t)}e^{\bar\eta\chi(t)}] =
e^{\eta\mbar\chi(s)}e^{\bar\eta\chi(s)}\, e^{2 \eta\bar\eta\,\mathfrak{n}_0
  (t-s)}.\] It indicates that the increments $\chi(t+s)-\chi(s)$ are independent
and identically distributed.

Note that the above limit is not the same as the limit $\kappa\to 0$ but with
$D\equiv \kappa\mathfrak{n}_0$ fixed which would gives
\begin{eqnarray*}
[\, \xi(t),\mbar\xi(s)\,]=0,\quad
\mathbb{E}[\mbar\xi(t)\xi(s)]=2D\,e^{i\nu(t-s)}\,\text{min}(t,s)
\end{eqnarray*}
and conditional expectations
\[ \mathbb{E}_s[ e^{\mu\mbar \xi(t)}e^{\bar\mu \xi(t)}] = e^{\mu e^{i\nu
    (t-s)}\mbar\xi(s)}e^{\bar\mu e^{-i\nu (t-s)}\xi(s)}\, e^{2 \mu\bar\mu\,D
  (t-s)},\] for $t>s$. It describes two commuting Brownian motions. The system
algebra is still that of the harmonic oscillator, $[z,\mbar z]=1$. The flow
simply corresponds to add the random $\mathbb{C}$-number $\xi(t)$ to the canonical
operators $z$, an operation which obviously preserves the commutation relations.
The Linbladian also simplifies in this limit, since only the double commutator
remains.

\subsection{Semi-classical limit} 
To take the semi-classical limit we put back the $\hbar$ factors and set
$\epsilon=\hbar\omega_0$ and replace $\varphi_\alpha$ by $\hbar\varphi_\alpha$
so that $\varphi_\alpha$ has the dimension of a frequency. The hamiltonian is
then:
\[ H= \sum_\alpha \hbar \omega_\alpha \mbar a_\alpha a_\alpha +
\lambda\hbar\big(\mbar w \sum_\alpha \varphi_\alpha a_\alpha +w \sum_\alpha \bar
\varphi_\alpha \mbar a_\alpha\big) + \hbar\omega_0\, \mbar w w. \] 
The large time limit is taken in the same way. 
The factors $\hbar$ compensate in evaluating
$z(t)$ so that we again have $z(t)=e^{-\bar\gamma t}z +\xi(t)$ with
$[\xi(t),\mbar \xi(s)]=G(t,s)$ where
\[ G(t,s)= e^{-\bar\gamma t-\gamma s}\,
(e^{(\gamma+\bar\gamma)\text{min}(t,s)}-1)\]

The two point function is still given by $\mathbb{E}[
\mbar\xi(t)\xi(s)]=\mathfrak{n_0}G(s,t)$.  The occupation number
$\mathfrak{n}_0$ may be estimated as if it is given by thermal equilibrium at
some temperature $T$, that is $\mathfrak{n}_0=1/(e^{\hbar\omega_0/T}-1)$ and
$\mathfrak{n}_0\simeq T/\hbar\omega_0$ as $\hbar$ goes to $0$.

The semi-classical limit is thus $\hbar\to 0$ with $\mathfrak{n}_0\hbar\equiv
T/\omega_0$ fixed.

Set $\gamma=\kappa+i\nu$ and let $X(t)=\sqrt{\hbar}\, \re z(t)e^{-i\nu t}$ and
$Y(t)=\sqrt{\hbar}\, \im z(t)e^{-i\nu t}$. Then, in the semi-classical limit,
$X(t)$ and $Y(t)$ are two commuting gaussian processes with~\footnote{With our
  convention, $X$ has dimension of $\sqrt{\hbar}$ since $X\simeq\sqrt{\hbar}\,
  \re z$. To go to the physical picture with $X$ having dimension of a length, one
  would have to introduce the mass scale $m$ of the harmonic oscillator and set
  $X\simeq \sqrt{\hbar/m\omega_0}\,\re z$.}
\[ \mathbb{E}[\, X(t)X(s)\,]= \mathbb{E}[\, Y(t)Y(s)\,]=\frac{T}{2\omega_0}\,
e^{-\kappa(t+s)}\big(e^{2\kappa\,\text{min}(t,s)}-1\big).\] They are two
commuting Ornstein-Uhlenbeck processes. As $\kappa\to0$, $X(t)/\sqrt{\kappa}$
and $Y(t)/\sqrt{\kappa}$ goes over two (un-normalized) commuting Brownian
motions.

\subsection{Decoherence and thermalisation}
Let us first consider the evolution of a density matrix which initially is
simply the pure state $\rho_0=|\alpha\ket\bra\alpha|$ with $|\alpha\ket$ the
coherent state $|\alpha\ket\equiv e^{-\alpha\bar\alpha/2}\, e^{\alpha \mbar
  z}|0\ket$. At time $t$, the density matrix is determined by duality via
$\text{Tr}_{\Gamma_S}(\rho_t\, a)=\text{Tr}_{\Gamma_S}(\rho_0\,\Phi_t(a))$ with
$\Phi_t$ given in eq.(\ref{stomap}). Using $\text{Tr}_{\Gamma_S}(\rho_0\,
e^{\mu\mbar z}e^{\bar\mu z})=e^{\alpha\bar\mu+\bar\alpha \mu}$, this gives
\[ \text{Tr}_{\Gamma_S}(\rho_t\, e^{\mu\mbar z}e^{\bar\mu z})=
e^{\alpha\bar\mu(t)+\bar\alpha \mu(t)}\,
e^{\mathfrak{n}_0(\mu\bar\mu-\mu(t)\bar\mu(t))}.\] This is solved in terms of
superposition of coherent states by:
\[ \rho_t = \int \frac{dad\bar a}{\pi \sigma_t}\ |a\ket\,
e^{-\frac{1}{\sigma_t}(a-\alpha(t))(\bar a - \bar \alpha(t))}\, \bra a|.\] with
$\sigma_t=\mathfrak{n}_0(1-e^{-2\kappa t})$ and $\alpha(t)=e^{-\bar\gamma
  t}\alpha$, an evolution equation similar to that of $z$ but without the
quantum noice. The interpretation is clear: the density matrix is an
(incoherent) sum of coherent states centred around the un-noisy solution
$\alpha(t)$ with dispersion $\sigma_t$ as in the classical case. As it should
be, it converges at large time towards the thermal invariant measure, since:
\[ \rho_\infty = \int \frac{dad\bar a}{\pi \mathfrak{n}_0}\ |a\ket\,
e^{-\frac{a\bar a}{\mathfrak{n}_0}}\, \bra a| = \frac{1}{\mathfrak{n}_0+1}\ e^{-\sigma\,
  \mbar z z},\] with $e^\sigma=(\mathfrak{n}_0+1)/\mathfrak{n}_0$. The
thermalisation time is
\[ \tau_\text{therm}= 1/\kappa.\]

Let us now consider an initial density matrix which is still that of a pure
state but for a state linear combination of coherent states, a kind of
Schrodinger cat:
\[\rho_0=\big(u |\alpha\ket+v|\beta\ket\big)\big(\bar u \bra\alpha| + \bar v \bra\beta|\big),\]
with $|\alpha-\beta|$ large. We explicitely write $\rho_0 = u\bar u\,
\rho^{\alpha\alpha}_0 + u\bar v\, \rho^{\alpha\beta}_0 + v\bar u\,
\rho^{\beta\alpha}_0 + v\bar v\, \rho^{\beta\beta}_0 .$ By linearity of the
evolution equation, the density matrix at time $t$ will be of form
\[ \rho_t = u\bar u\, \rho^{\alpha\alpha}_t + u\bar v\, \rho^{\alpha\beta}_t +
v\bar u\, \rho^{\beta\alpha}_t + v\bar v\, \rho^{\beta\beta}_t.\] The evolution
of each of the terms is determined as above. For instance:
\[ \rho_t^{\beta\alpha} =\bra\alpha|\beta\ket\, \int \frac{dad\bar a}{\pi
  \sigma_t}\ |a\ket\, e^{-\frac{1}{\sigma_t}(a-\beta(t))(\bar a - \bar
  \alpha(t))}\, \bra a|,\] or equivalently,
\[ \bra b|\rho_t^{\beta\alpha}|c\ket= \bra\alpha|\beta\ket \bra b|c\ket\,
e^{-\frac{1}{\sigma_t+1}(c-\beta(t))(\bar b-\bar\alpha(t))},\] 
with $|b\ket$ and $|c\ket$ coherent states. In order to code
for the decoherence, let us consider the ratio of the matrix elements of the
'mixed' density matrices, $\rho_t^{\beta\alpha}$ or $\rho_t^{\alpha\beta}$, over
the pure ones, $\rho_t^{\alpha\alpha}$ or $\rho_t^{\beta\beta}$, defined by:
\[ \mathfrak{R}_t= \frac{\bra b|\rho_t^{\beta\alpha}|c\ket\,\bra
  b|\rho_t^{\alpha\beta}|c\ket}{\bra b|\rho_t^{\alpha\alpha}|c\ket\,\bra
  b|\rho_t^{\beta\beta}|c\ket},\] with $\mathfrak{R}_0=1$.  An explicit
computation shows that this ratio is independent of $b$ and $c$ and given by
\[ \mathfrak{R}_t = |\bra\alpha|\beta\ket|^2\
|\bra\alpha(t)|\beta(t)\ket|^{-2/(\sigma_t+1)}.\]
Recall that $|\bra\alpha|\beta\ket|^2=e^{-|\alpha-\beta|^2}$.
At short time, $\sigma_t\simeq 2\mathfrak{n}_0\kappa\, t$, so that 
\[ \log \mathfrak{R}_t \simeq -2(\mathfrak{n}_0+1)\kappa\, t\,
|\alpha-\beta|^2,\] at short time. Hence, the mixed density matrices decrease
exponentially faster than the pure ones.
 
Thus there are two regimes: first a short decoherence time
$\tau_\text{coher}=\frac{1}{\kappa(\mathfrak{n}_0+1)|\alpha-\beta|^2}$, whose
semi-classical limit is (recall that $\mathfrak{n}_0\simeq T/\hbar\omega_0$ and
$\delta X\simeq \sqrt{\hbar}\, \delta\alpha$)
\[ \tau_\text{coher}\simeq \frac{\omega_0}{\kappa T}\, \frac{\hbar^2}{|\delta
  X|^2},\] and which decreases as the inverse power of the square distance
between the states, and then a longer thermalisation time
$\tau_\text{therm}\simeq 1/\kappa$.

\section{Conclusion}

Motivated by an initial problem related to the quantum Hall effect, these notes
have investigated a simple open quantum system. Starting from a traditional
coupled system-reservoir model, we were able to study most of its salient
features in the weak-coupling/long-time limit, from the more physical to the
more mathematical.

The model is simple enough to be solved by several methods. One of these
concentrates on the limiting behavior of expectations and commutators, without
paying attention to the limit of operators themselves. The output is an abstract
(almost axiomatic) representation result for the limiting process in Fock
spaces. Another approach, which leads to the same result, pays more attention to
the relationship between the bath oscillators of the initial model and their
weak-coupling/long-time cousins. As expected, dissipation
arises via destructive interferences. 

Along the way, we retrieve a number of well-studied objects that we connect
together, like the damped harmonic oscillator on the physical side, or the
quantum stochastic calculus and quantum geometry on the mathematical side. We
pay a particular attention to the explicit construction, in this simple case, of
a quantum filtration and the associated quantum conditional expectations. 
This leads to the specification of the (mathematical) dilation of the quantum  dynamical maps. It
gives access to multi-time correlation functions that cannot be computed via the
knowledge of the system time dependent density matrix only. 
The quantum phase space Brownian motion we obtain possesses 
peculiar properties analogous to those of the classical one.
We observe the remarkable fact that despite non-trivial non-commutative effects, the
conditional expectations have the property to be transparent to
multiplication by measurable observables (on the left or on the right). 
We also do a sample computation of decoherence.

One of the missing probabilistic concepts is that of stopping time -- 
which is instrumental for computing exit probabilities or harmonic measures. The
concept is defined in the mathematical literature, but the examples are few and
a bit artificial. In the present case, and if the system starts in
the ground state, one can define the time at which the system leaves the ground
state. This can be checked to be a stopping time, i.e. deciding if it occurs
before time $t$ can be done by looking only at conditional expectations up to
time $t$. Alas, due to continuous time, this time vanishes almost surely.
However, our discussion of filtration and conditional expectations can be extended straightforwardly to a discrete version of eq.(\ref{qdiif}), namely $z_0=z$ and $
z_{n}=\lambda z_{n-1} +(1-|\lambda|^2)^{{1}/{2}}\, a_n$ for $n\geq 1$ 
(where $|\lambda| \in [0,1[$  and the $a_n$'s are standard independent annihilation operators).
Then one checks without difficulty that, starting the
system in the ground state at $n=0$, the time at which it leaves the ground
state is a stopping time, which follows a geometric distribution. Other examples
of meaningful stopping times, in the discrete and especially in the continuous
time case, would surely be desirable.

We hope to come back in a near future to all these themes in other simple open quantum systems -- including generalisations to finite dimensional quantum systems and the associated quantum stochastic processes, aspects of quantum geometry of quantum phase space Brownian motion --  and to the more puzzling case of dissipative effects in the quantum Hall effect and their relations to the behaviour of the first Landau level coupled to a reservoir.
\vfill \eject

\section{Appendix: Algebraisation of stochastic processes}
\label{App:stochastic} The aim of this appendix is to recap the algebraic
structure underlying classical stochastic processes for comparison with the
quantum theory. Although we did not find it gathered in one reference,
we have no doubt that the material presented here is already somewhere
(dispersed) in the literature.

Suppose we are given an adapted stochastic process, defined on a probability
space $\Omega$, with measure $\mathbb{E}$ and filtration ${\cal F}_t$, and
valued in some measured space $V$ with measure $dy$. For $\omega\in\Omega$, let
$t\to X_t(\omega)$, or simply $X_t$, be the realisation of the process. By
construction this defines a path on $V$ (for simplicity we may assume that it is
continuous). The measure $\mathbb{E}$ then induces a measure on paths on $V$.

To simplify the writing, we shall restrict ourselves to time-homogeneous Markov
processes but most of the algebraic structure we shall describe remains valid
for non Markovian processes (up to simple reformulations). Let $P_t(dy,x)$
be the transition kernel, i.e. the probability density for the proccess started
at $x$ at time $t=0$ to be in the neighbourhood of $y$ at time $t$, so that for
any (good enough) function $f$ from $V$ to, say, $\mathbb{C}$ we have:
\[ \mathbb{E}_x[\, f(X_t)] = \int f(y)\, P_t(dy,x) .\]

Let ${\cal A}_0$ be the algebra of measurable functions on $V$ with values, say,
in $\mathbb{C}$ (with the Borel $\sigma$-algebra), ${\cal A}_t$ be the algebra
of ${\cal F}_t$-measurable functions and ${\cal A}_\infty$ some inductive limit
of these algebras adapted to the filtration: ${\cal A}_0\equiv\text{Func}(V\to
\mathbb{C})$ and
\[ {\cal A}_t\equiv \{ f\in\text{Func}(\Omega\to \mathbb{C}),\ \text{s.t.}\ f\
\text{is}\ {\cal F}_t\text{-measurable}\}.\] Since to any $\omega\in \Omega$ we
associate a path $t\to X_t(\omega)$ on $V$, function of ${\cal A}_t$ are
functions on parameterized paths $[0,t]\to V$ of "length" $t$ drawn on $V$. The
algebra ${\cal A}_\infty$ is the algebra of $\mathbb{E}$-measurable functions.
We identify ${\cal A}_0$ with ${\cal A}_{t=0}$. Of course, since ${\cal F}_t$ is
a filtration, we have the embeddings:
\[ {\cal A}_0\subset {\cal A}_s\subset {\cal A}_t \subset {\cal A}_\infty ,\quad
\text{for}\ s<t.\] 
Naively, ${\cal F}_s$-measurable functions are functions only
of the portion of the path $t\to X_t(\omega)$ for $t\in[0,s]$, i.e. they code
about the knowledge of the process from time $0$ to time $s$.

The data of the stochastic flow $X_t$ (tautologically) induces a map $J_t$ from
${\cal A}_0$ to ${\cal A}_t$ which to any test function $f$ on $V$ associate an
${\cal F}_t$-measurable function by:
\[ J_t:\ {\cal A}_0 \to {\cal A}_t,\quad 
J_t(f)(\omega)=f(X_t(\omega)).\]
Given $P_t$, we may define the stochastic map $\Phi_t$ on ${\cal A}_0$ by:
\[ \Phi_t(f)(x)\equiv \mathbb{E}_x[\, f(X_t)\,],\quad f\in{\cal A}_0.\]
Alternatively, $\Phi_t= \mathbb{E}\circ J_t$, with a slight abuse of notation.
As $P_t$, they define a semi-group $\Phi_t\circ\Phi_s=\Phi_{t+s}$. Note that
$\Phi_t$ is defined using only one-point functions (involving only one time).

Conditional expectations are maps from ${\cal A}_t$ onto ${\cal A}_s$ for $s<t$:
\[ \mathbb{E}[\, F_t\,|{\cal F}_s]\in{\cal A}_s,\quad \text{for}\ F_t\in{\cal
  A}_t.\] They are neutral with respect to multiplication by ${\cal
  F}_s$-measurable functions:
$$\mathbb{E}[\, F_s\, F\, |{\cal F}_s]= F_s\, \mathbb{E}[\, F\, |{\cal F}_s].$$
As a consequence, they are nested projectors since $\mathbb{E}[F_s|{\cal
  F}_s]=F_s$ for $F_s\in{\cal A}_s$ and
\[ \mathbb{E}[\, \mathbb{E}[\, F\, |{\cal F}_s]\, |{\cal F}_t ]=\mathbb{E}[\,
F\, |{\cal F}_{\text{min}(s;t)}],\] for any $\mathbb{E}$-measurable function
$F$.

The data of $P_t(dy,x)$ specifies the process. In particular, by conditioning
recursively on the position of the path at intermediate times, multipoint
correlation functions may be evaluated as
\begin{eqnarray*}
  &&\mathbb{E}_x[\, f_1(X_{t_1})\cdots f_n(X_{t_n})\, ] \\
  &&~~~~~~=\int  f_1(y_1)P_{t_1}(dy_1,x)\cdots \int
  f_n(y_n)P_{t_n-t_{n-1}}(dy_n,y_{n-1}), 
 \end{eqnarray*}
 for $t_1<\cdots<t_n$.  By definition conditioning on $X_{t=0}=x$ amounts to
 put the Dirac point measure on ${\cal F}_0$, so that these multipoint expectations may also be
 written in terms of the maps $\Phi_t$ as:
\begin{eqnarray*}
  \mathbb{E}[\, f_1(X_{t_1}) \cdots f_n(X_{t_n}) |{\cal F}_0\, ] 
  = \Phi_{t_1}\big(f_1\, \Phi_{t_2-t_1}\big(\cdots f_{n-1}\,
  \Phi_{t_n-t_{n-1}}(f_n)\big)\big) 
 \end{eqnarray*}

 More intrinsically (and for non Markovian processes), using the nested property
 of conditional expectations, for any ${\cal F}_{t_j}$-measurable functions
 $F_j$, we can recursively write
\begin{eqnarray*}
  \mathbb{E}[\, F_1 \cdots F_n | {\cal F}_{0} ] 
  &=&\mathbb{E}[\,F_1 \cdots F_{n-1} \mathbb{E}[F_n|{\cal F}_{t_{n-1}} ]|{\cal
    F}_{0} \, ] \\ 
  &=&\mathbb{E}[\,F_1\mathbb{E}[F_{2}\cdots \mathbb{E}[F_n|{\cal
    F}_{t_{n-1}}]\cdots|{\cal F}_{t_{2}}]|{\cal F}_{0} \, ]  
\end{eqnarray*}
We see that this formula can be written in terms of maps $\Phi_{t;s}$, which map
${\cal F}_t$-measurable functions onto ${\cal F}_s$-measurable functions,
defined for $s<t$ by:
\[ \Phi_{t;s}(F)\equiv \mathbb{E}[\,F\,|{\cal F}_s]\in {\cal A}_s,\quad \text{
  for}\ F\in{\cal A}_t.\] Indeed,
\begin{eqnarray*}
  \mathbb{E}[\, F_1 \cdots F_n |{\cal F}_{0} \, ] 
  =\Phi_{t_1,0}\big(F_1\, \Phi_{t_2;t_1}\big(\cdots F_{n-1}\,
  \Phi_{t_n;t_{n-1}}(F_n)\big)\big) 
 \end{eqnarray*}
 For $F_j=f_j(X_{t_j})$ and time homogeneous Markov processes, this last formula coincides
 with the previous one since then $\Phi_{t;s}(f(X_s))=\Phi_{t-s}(f)$ for $s<t$.

 \section{Appendix: Another approach to the long time limit} \label{App:weak}
 The aim of this Appendix is to re-derive the
 long-time/weak-coupling/continuous-density limit of Section \ref{Sec:wlim}
 differently in way which makes more transparent the fact that damping arises
 from of an infinite number of destructive interferences.

 Without loss of generality, we assume that the $\varphi_\alpha$'s are non
vanishing. We also assume that the $\omega_\alpha$'s are all distinct.

Define a rational function 
\[S(\Omega)\equiv \Omega-\epsilon - \lambda^2
\sum_{\alpha}\frac{|\varphi_\alpha|^2}{\Omega-\omega_\alpha}.\] 
Its degree is equal to the number of oscillators in the bath plus one.  Direct
substitution shows that $\Omega$ is an eigen-frequency of the coupled system (i.e.
the coupled equations have a solution for which both $w$ and the $a_\alpha$'s
have a time dependence $e^{-i\Omega t}$) if and only if $S(\Omega)=0$.  The
variations of $S$ show immediately that the solutions are real and intertwine
the $\omega_\alpha$'s: as expected, there is one more solution than the number
of oscillators in the bath. In particular, there is at most one negative
eigenfrequency. Let $\{\Omega_k\}$ be the set of eigenfrequencies of the coupled
system. 

\begin{figure}
\center{\includegraphics[width=0.8\textwidth]{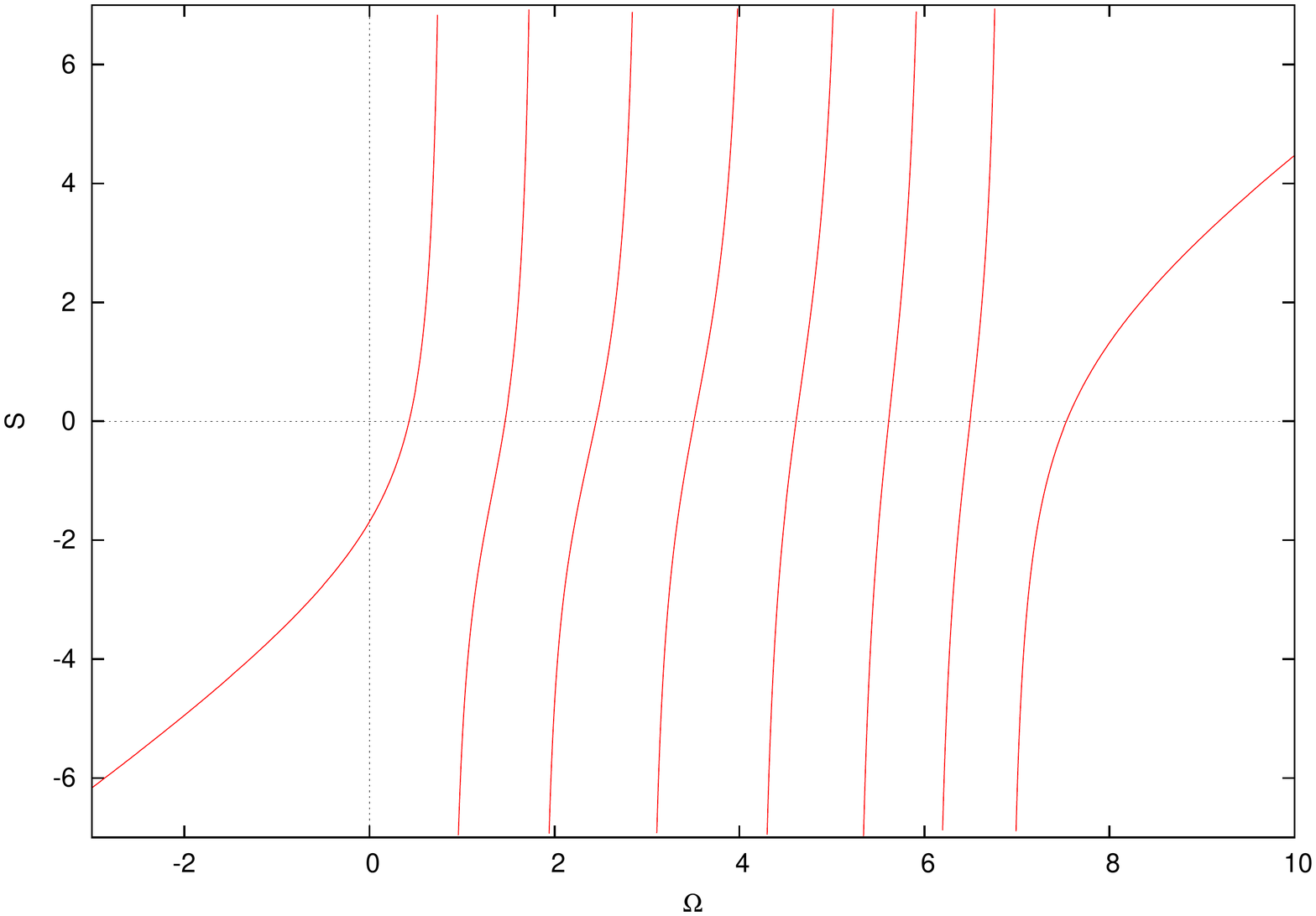}}
\caption{\emph{The typical shape of the function $S(\Omega)$.}}
\label{fig:S}
\end{figure}

The Hamiltonian is bounded below if and only no solution is negative, i.e. if
\[\epsilon > \sum_{\alpha}\frac{|\varphi_\alpha|^2}{\omega_\alpha}.\]   
We assume that this condition is satisfied in what follows. 

Writing
\[\frac{1}{S(\Omega)}=\sum_k \frac{R_k}{\Omega-\Omega_k},\] 
we claim that the general solution of the coupled system is
\begin{eqnarray*}
  w(t) & = & \sum_k R_k e^{-i\Omega_k t}\left( w(0) + \sum_{\alpha} a_\alpha(0)
    \frac{\lambda \varphi_\alpha}{\Omega_k-\omega_\alpha}\right) \\ 
  a_\beta(t) & = & \sum_k R_k e^{-i\Omega_k t}\left( w(0) + \sum_{\alpha}
    a_\alpha(0) 
    \frac{\lambda \varphi_\alpha}{\Omega_k-\omega_\alpha}\right)\frac{\lambda
    \bar{\varphi}_\beta}{\Omega_k-\omega 
    _\beta}. \end{eqnarray*} 
These formul\ae\ could be obtained mechanically via residue calculus using the
inverse Laplace transform solution presented in Section \ref{Sec:wlim}. 
But a direct argument is easy too.  
For fixed $k$, the coefficient for $a_\beta(0)$ is just $\frac{\lambda
  \bar{\varphi}_\beta}{\Omega_k-\omega_\beta}$ times the coefficient for $w$, so
this is 
indeed a family of solutions of the coupled equations. What remains to be
checked is that the initial conditions are satisfied. This is the case because
\begin{eqnarray*}
\sum_k R_k & = & 1, \\  \sum_k \frac{R_k}{\Omega_k-\omega 
    _\alpha} & = & 0 \text{ for any } \alpha , \\
\sum_k \frac{R_k}{(\Omega_k-\omega 
    _\alpha)(\Omega_k-\omega 
    _\beta)} & = & \delta_{\alpha ;\beta}\,\frac{1}{\lambda^2 \bar{\varphi}_\beta
    \varphi_\alpha}  \text{ for any } \alpha,\beta.
\end{eqnarray*}
The first equation expresses that $S(\Omega)$ behaves like $\Omega$ for large
$\Omega$, so that $1/S(\Omega)$ behaves like $1/\Omega$. The second equation
expresses that $S(\Omega)$ has a pole at each $\omega_{\alpha}$ so that
$1/S(\omega_\alpha)=0$. For the third identity, we consider two cases. If
$\alpha\not=\beta$, writing $\frac{1}{(\Omega_k-\omega
  _\alpha)(\Omega_k-\omega _\beta)}= \frac{1}{\omega
  _\alpha-\omega _\beta}\left(\frac{1}{\Omega_k -\omega_\alpha}- \frac{1}{\Omega
  _k-\omega _\beta}\right)$, the third identity is a consequence of the
second. For $\alpha=\beta$, the left-hand side is nothing but
$-\frac{d}{d\Omega}\frac{1}{S(\Omega)}$ evaluated at $\Omega=\omega_\alpha$. As
$S(\Omega)\sim \frac{-\lambda^2 |\varphi_\alpha|^2}{\Omega-\omega_\alpha}$ for
$\Omega$ close to $\omega_\alpha$ the result follows.

 ~From now on, our aim is to compute the behavior of the $\Omega_k$'s and $R_k$'s
when the $\omega_\alpha$'s and $\varphi_\alpha$'s are adjusted to mimic a
smooth density. We choose a smooth positive density function $r(\omega)$
and consider Riemman sums with a mesh $\eta$, a small parameter with the
dimension of a frequency, leading to
\[ S(\Omega)=\Omega-\epsilon-\eta \sum_{l\in [1,L]} \frac{\lambda^2 r(\eta l
  )}{\Omega-\eta l}.\]
Later, we can let $\eta\rightarrow 0^+$ and $L\rightarrow +\infty$. 
We assume that $\epsilon > \sum_{l\in [1,L]} \frac{r(\eta l)}{l}$, which is
  ensured for $\eta$ small enough if $\epsilon > \int_0^{+\infty}
  r(\omega)\frac{d\omega}{\omega}$, an integral that we assume to be convergent.
The zeroes of $S$ can be ordered as $0 < \Omega_0 < \eta < \Omega_1 < 2\eta
< \cdots < \Omega_{L-1} < L\eta < \Omega_L$. Though not stricly evenly spaced,
the $\Omega_k$'s are good enough to define Riemman sums.  

The phases of the $\varphi_\alpha$'s can be reabsorbed in the
bath oscillators, so we may assume that the $\varphi_\alpha$'s are real and
positive. Then, we can write the coupled time evolution as 
\begin{eqnarray*} w(t) & = & \sum_{k\in [0,L]} R_k e^{-i\Omega_k t}\left( w(0) +
    \sum_{l\in [1,L]} a_l(0) \frac{\lambda \sqrt{\eta r(\eta l)}}{\Omega_k-\eta
      l}\right) \\
  a_{l'}(t) & = & \sum_{k\in [0,L]} R_k e^{-i\Omega_k t}\left( w(0) + \sum_{l\in
      [1,L]} a_l(0) \frac{\lambda \sqrt{\eta r(\eta l)}}{\Omega_k-\eta l}\right)
  \frac{\lambda \sqrt{\eta r(\eta l')}}{\Omega_k-\eta l'}.
\end{eqnarray*}

Fix $\Omega < \eta L $ and take $k$ such that $k\eta \leq \Omega < (k+1)\eta $.
We want to find an approximate formula for $S(\Omega)$ for fixed
$\Omega$ when $\eta$ is small and $L$ large. We shall see that the condition
$\eta \log L$ small is important. 

More precisely, we let $\eta\to0$ but write $\Omega=\eta(k +\delta)$, $\delta \in ]0,1[$, 
so that the interval between two adjacent eigenvalues is blown up to size $1$.
We need to compute
$\eta(k +\delta)-\epsilon -\lambda^2 \sum_{l\in [1,L]} \frac{
  r(\eta l )}{k -l+\delta}=0$. We split the sum as
\[ \sum_{l\in [1,L]} \frac{r(\eta l )}{k -l+\delta} =
\frac{r(\eta k )}{\delta}+\sum_{l\in [1,L],l\neq
  k}\left(\frac{r(\eta l )}{k -l+\delta}-\frac{r(\eta l )}{k
    -l}\right) +\sum_{l\in [1,L],l\neq k}\frac{r(\eta l )}{k -l}.\] The
last sum is well approximated by $$\hat{r}(\Omega)\equiv \fint_0^{+\infty} d\omega \frac{
  r(\omega)}{\Omega-\omega}$$ up to $O(\eta)$. In the remaining sum, writing
$r(\eta l)=r(\eta k)+(r(\eta l)-r(\eta k))$, the second term is $O(\eta \log
L)$. When these errors can be neglected, a good approximation for $S(\Omega)$
is
\[S(\Omega)=\Omega -\epsilon -\lambda^2 \left( r(\Omega)\frac{\pi}{\tan \pi
    \delta}+\fint_0^{+\infty} d\omega \frac{r(\omega)}{\Omega-\omega}\right).\]
This equation gives an approximation for the $\Omega_k$'s close to
$\Omega$ by choosing $\delta=\delta_k$ such that the right-hand side
vanishes. This is not directly needed, but it is the crucial ingredient to control 
$R_k$ in this region. Indeed, to get $S'(\Omega)$ we can take the derivative
with respect to $\delta$ and divide by $\eta$. This leads to 
$\eta S'(\Omega)=\lambda^2 \pi^2 r(\Omega)\left(1+\frac{1}{\tan^2 \pi
    \delta}\right)$. But $S'(\Omega_k)=1/R_k$, leading to
\[R_k=\frac{\eta r(\Omega)}{\lambda^2\left( \pi^2 r(\Omega)^2
    +\left(\frac{\Omega-\epsilon}{\lambda^2}-\fint_0^{+\infty} d\omega \frac{
        r(\omega)}{\Omega-\omega}\right)^2 \right)}\equiv \eta R(\Omega)\] for
$k\eta$ close to $\Omega$. So indeed, in this approximation $R_k$ is obtained by
discretizing a smooth function.  For instance,
\[ \sum_{k\in [0,L]} R_k e^{-i\Omega_k t} \sim  \int_0^{+\infty}d\Omega
\, e^{-i\Omega t} R(\Omega).\]
This is enough to treat the term proportional to $w(0)$ in $w(t)$. To deal with
the oscillator contribution in $w(t)$, we need to control the sum
\[
\sum_{k\in [0,L]} R_k e^{-i\Omega_k t} \frac{1}{\Omega_k-\eta l}
\]
By contruction, at $t=0$ this sum vanishes, and we may subtract $0=e^{-i\eta l  t}\sum_{k\in
  [0,L]} R_k  \frac{1}{\Omega_k-\eta l}$ to get 
\[
\sum_{k\in [0,L]} R_k e^{-i\Omega_k t} \frac{1}{\Omega_k-\eta l} \sim
\int_0^{+\infty}d\Omega \frac{e^{-i\Omega t}- e^{-i\eta l t}}{\Omega-\eta l}
R(\Omega).
\]
If we define $a_l=\sqrt{\eta}a(\eta l)$ then in the continuum limit we get
$[a(\omega),\mbar{a}(\omega')]=\delta(\omega-\omega')$, and we find 
\[w(t)=w\int_0^{+\infty}d\Omega \, e^{-i\Omega t} R(\Omega)
+\lambda \int_0^{+\infty}d\Omega d\omega\, \frac{e^{-i\Omega t}- e^{-i\omega
    t}}{\Omega-\omega}a(\omega)\sqrt{r(\omega)}R(\Omega) .\]

The limit
$\lambda \rightarrow 0^+$ in $z(t)=w(t/\lambda^2)e^{i\epsilon t/\lambda^2}$ is
now straigthforward. We set $\Omega \rightarrow \epsilon +\lambda ^2 \Omega$,
$\omega \rightarrow \epsilon +\lambda ^2 \omega$. Observe that $\lambda
a(\epsilon +\lambda ^2 \omega)$ is still a normalized oscillator, which we
denote by $a_\omega$. Noting that $R(\epsilon +\lambda
^2 \Omega)\sim \frac{r(\epsilon)}{\lambda ^2 \left(\pi^2 r(\epsilon)^2+
    (\Omega-\hat{r}(\epsilon))^2\right)}$, which we
denote by $R_\Omega$, we find that
 \begin{eqnarray*} z(t) & = & z \int_{-\infty}^{+\infty}d\Omega \, e^{-i\Omega t}
   R_\Omega +\int_{-\infty}^{+\infty}d\Omega d\omega\, \frac{e^{-i\Omega t}-
     e^{-i\omega
       t}}{\Omega-\omega}a_\omega\sqrt{r(\epsilon)}R_\Omega \\
   & = & z\int_{-\infty}^{+\infty}d\Omega e^{-i\Omega
     t}\frac{r(\epsilon)}{\pi^2 r(\epsilon)^2+ (\Omega-\hat{r}(\epsilon))^2} \\ &
   + & \int_{-\infty}^{+\infty} d\Omega d\omega\, \frac{e^{-i\Omega t}-
     e^{-i\omega t}}{\Omega-\omega}a_\omega\frac{r(\epsilon)^{3/2}}{\pi^2
     r(\epsilon)^2+ (\Omega-\hat{r}(\epsilon))^2}. \end{eqnarray*} 

Integration
over $\Omega$ is responsible for destructive interferences.  Performing this
integral and using that $\bar{\gamma}\equiv\pi r(\epsilon)+ i\hat{r}(\epsilon)$
leads to
 \begin{equation}\label{ztmich}
 z(t)=z(0)e^{-\bar{\gamma}t}+\sqrt{\frac{\Re e \,\bar{\gamma}}{\pi}}
  \int_{-\infty}^{+\infty}d\omega\, a_\omega\frac{e^{-i\omega
      t}-e^{-\bar{\gamma}t}}{\omega+i\bar{\gamma}}. 
 \end{equation}
 This gives the spectral representation of $z(t)=ze^{-\bar{\gamma}t}+\xi(t)$. 
 One checks that $[\xi(t),\mbar\xi(s)]=G(t,s)$ as should be.

The spectral representation also allows to check easily that the coupled
bath-reservoir system is still hamiltonian, with Hamiltonian
\begin{eqnarray*}
 H= \int d \omega \,  \omega \mbar a_\omega a_\omega + \sqrt{\frac{\Re e
     \,\bar{\gamma}}{\pi}} \int d \omega \,
 \left(\mbar z a_\omega + z  \mbar a_\omega \right) + \Im m  \,\bar{\gamma}
 \mbar z z 
\end{eqnarray*} 
Checking that this leads to the spectral representation is done routinely via
the Laplace transform, interpreting at some point $\int d \omega \,
\frac{1}{p+i\omega}$ as $\fint d \omega \, \frac{1}{p+i\omega} =\pi$ for $\Re e
\, p > 0$. Note that the hamiltonian is not bounded below due to the appearance
of negative frequencies. This is simply due to the fact that taking the long
time limit leads to shift the origin of energies by $\epsilon$. 

The relation between the coupling constant $\sqrt{\frac{\Re e\,\bar{\gamma}}{\pi}}$ 
and the friction coefficient $\Re e \,\bar{\gamma}$ is no surprise 
(though it is usually interpreted as the Einstein relation only
after introduction of a temperature).

To diagonalize the dynamics, define (recall that $R_\Omega=r(\epsilon)/|\Omega+i\bar\gamma|^2$)
\[b_\Omega\equiv R_\Omega^{1/2} z
+\left(\frac{r(\epsilon)}{R_\Omega}\right)^{1/2}\left(
 \frac{a_\Omega}{\Omega+i\bar{\gamma}}-R_\Omega 
\int_{-\infty}^{+\infty}d\omega\, \frac{a_\omega}{\omega+i\bar{\gamma}} \right).\]
One checks that the $b$'s are normalized oscillators
\[ [b_\Omega, \mbar{ b_{\Omega'}}]=\delta(\Omega-\Omega'),\]
such that $[H,b_\Omega]=-\Omega\, b_\Omega$, and that 
\[z(t)=\int_{-\infty}^{+\infty}d\Omega\, e^{-i\Omega t}\,R_\Omega^{1/2}\,
b_{\Omega}.\] 
Just as $z=\int_{-\infty}^{+\infty}d\Omega\, R_\Omega^{1/2}
b_{\Omega}$ commutes with the operators $a_\omega$ and $\mbar a_\omega$, one checks
that $\int_{-\infty}^{+\infty}d\omega\, \frac
{r(\epsilon)^{1/2}}{\omega+i\bar{\gamma}} a_{\omega}$ is a normalized
anihilation operator that commutes with the operators $b_\Omega$ and $\mbar b_\Omega$.

\section{Appendix: Proofs} \label{App:proof} Here we present the computation
which leads to the formula eq.(\ref{defEs}) for the conditional expectations.
Eq.(\ref{EsJt}) defines $\mathbb{E}_s$ on single-time operator. We have to
define it on product of time ordered operators, that is:
\[ \mathbb{E}_s[\, e^{\mu_1 \mbar\xi(t_1)}e^{\bar\mu_1\xi(t_1)}\cdots e^{\mu_N
  \mbar\xi(t_N)}e^{\bar\mu_N\xi(t_N)}\,] \] with $t_1<\cdots<t_N$.  
We impose that $\mathbb{E}_s$ is neutral with respect to left
multiplication by elements of $A_s$, i.e. $\mathbb{E}_s[ab]=a\mathbb{E}_s[b]$
for $a\in A_s$.  As a consequence, it is enough to consider times bigger than
$s$, that is $s<t_1<\cdots<t_N$. We may then recursively compute the conditional
expectation by imposing the required condition $\mathbb{E}_{s_1}\circ
\mathbb{E}_{s_2} = \mathbb{E}_{\text{min}(s_1,s_2)}$. Indeed, inserting first
$\mathbb{E}_s=\mathbb{E}_{s}\circ \mathbb{E}_{t_{N-1}}$ for $s<t_{N-1}$ in the
above conditional expectation and using that all operators $e^{\mu_j
  \mbar\xi(t_j)}e^{\bar\mu_j\xi(t_j)}$ with $j<N-1$ are left neutral with
respect $\mathbb{E}_{t_{N-1}}$ (by construction), we obtain
\begin{eqnarray*} &&\mathbb{E}_s[ e^{\mu_1
\mbar\xi(t_1)}e^{\bar\mu_1\xi(t_1)}\cdots e^{\mu_N
\mbar\xi(t_N)}e^{\bar\mu_N\xi(t_N)}\,]\\ 
=&& \hskip -.5 truecm 
\mathbb{E}_s[\, e^{\mu_1 \mbar\xi(t_1)}e^{\bar\mu_1\xi(t_1)}\cdots e^{\mu_{N-1}
\mbar\xi(t_{N-1})}e^{\bar\mu_{N-1}\xi(t_{N-1})}\, \mathbb{E}_{t_{N-1}}[
e^{\mu_N \mbar\xi(t_N)}e^{\bar\mu_N\xi(t_N)}] \,] \\ 
=&& \hskip -.5 truecm
\mathbb{E}_s[\,e^{\mu_1 \mbar\xi(t_1)}e^{\bar\mu_1\xi(t_1)}\cdots e^{\mu_{N-1}
\mbar\xi(t_{N-1})}e^{\bar\mu_{N-1}\xi(t_{N-1})}\times \\ && \times
\,e^{\mu_N(t_{N;N-1})
\mbar\xi(t_{N-1})}e^{\bar\mu_N(t_{N;N-1})\xi(t_{N-1})}\,]\,
e^{\mathfrak{n}_0[\mu_N\bar\mu_N-\mu_N(t_{N;N-1})\bar\mu_N(t_{N;N-1})]}
 \end{eqnarray*} with $t_{N;N-1}=t_N-t_{N-1}$ and where in the last equality we
used the formula for conditional expectations of single-time operators. The key
point is that now the conditional expectation $\mathbb{E}_s$ only involves the
$N-1$ times $t_1<\cdots<t_{N-1}$, so that we can recursively apply this procedure to
fully compute the conditional expectation.  After
appropriate reordering of all terms, we get the formula (\ref{defEs}).

One now verifies that the conditional expectations (\ref{defEs}) satisfy
\[ \mathbb{E}_{t}\circ \mathbb{E}_{s} =\mathbb{E}_{\text{min}(s,t)}. \]
This is true by construction for $s\leq t$. The proof for $t<s$ requires a tiny computation. One first gets the factors $X_s^{(N)}$ and $Y_s^{(N)}$ by applying $\mathbb{E}_s$. Applying next $\mathbb{E}_t$ we get an extra contribution to $X$ coming from the difference between reordering the operators before and after the projection by $\mathbb{E}_t$ which is
\[\hat X_{t;s}\equiv \sum_{i<j}[\bar\mu_i(t_{i;s})\mu_j(t_{j;s})-\bar\mu_i(t_{i;t})\mu_j(t_{j;t})].\]
One checks that $X_s^{(N)}+\hat X_{t;s}=X_t^{(N)}$. Similarly, acting next with $\mathbb{E}_t$ we get an extra contribution to $Y$ which is
\[ \hat Y_{t;s}=\sum_{i,j}[\bar\mu_i(t_{i;s})\mu_j(t_{j;s})-\bar\mu_i(t_{i;t})\mu_j(t_{j;t})].\]
One also checks that $Y_s^{(N)}+\hat Y_{t;s}=Y_t^{(N)}$.

Lastly we prove that the conditional expectations are right neutral by measurable multiplication, i.e. $\mathbb{E}_s[ba]=\mathbb{E}_s[b]a$ for $a\in A_s$. It is enough to check it for $a_0=e^{\mu_0\mbar\xi(t_0)}e^{\bar\mu_0\xi(t_0)}$ and $b$ a product of operators as above. If $a_0$ multiplies this product of operators on the right, we first have to move it to the left by using the commutation relations, then to use the left neutrality and apply $\mathbb{E}_s$ and, finally, to move it back to the right using again the commutation relations. The difference between the commutators before and after having applied $\mathbb{E}_s$ is:
\begin{eqnarray*}
&&\sum_j\big( \bar\mu_j\mu_0[\xi(t_j),\mbar\xi(t_0)]+ \bar\mu_0\mu_j[\mbar \xi(t_j),\xi(t_0)]\big)\\
&&+ \sum_j\big( \bar\mu_0\mu_j(t_{j;s})[\xi(t_0),\mbar\xi(t_s)]+ \mu_0\bar\mu_j(t_{j;s})[\mbar \xi(t_0),\xi(t_s)]\big).
\end{eqnarray*}
This vanishes thanks to the relations $\mu_j[\xi(t_0),\mbar\xi(t_j)]=\mu_j(t_{j;s})[\xi(t_0),\mbar\xi(t_s)]$, and their complex conjugate, valid for $t_0<s<t_j$.


\begin{thebibliography}{}

\bibitem{CL} A.O. Caldeira and A.J. Leggett, "Influence of dissipation on quantum tunneling in macroscopic systems", Phys. Rev. Lett., 46, (1981) 211.

\bibitem{Gard-Zoller}  C.W. Gardiner and P. Zoller, {\it "Quantum Noise"}, 2nd edition, Springer, 2000.

\bibitem{Breuer-Pet} H.P. Breuer and F. Petruccione, {\it "The theory of open quantum systems"}, Oxford University Press, 2006.

\bibitem{Legget-review} A. Leggett, S. Chakravarty, A. Dorsey, M. Fisher, A. Garg, and W. Zwerger, {\it "Dynamics of the dissipative two-state system"}, Rev. Mod. Phys., 59, (1987) 1.
 
\bibitem{Davies73} E.B. Davies, "The harmonic oscillator in a heat bath", Commun. Math. Phys. 33 (1973) 171-186.

\bibitem{levy} P. Levy, {\it "Processus stochastiques et mouvement Brownien"}, Gauthier-Villars, Paris, 1965.

\bibitem{lawler} G. Lawler, {\it "Conformally invariant process in the plane"}, American Mathematical Society, 2005

\bibitem{hope to come} M. Bauer and D. Bernard, to appear.

\bibitem{landau} R.E. Prange and S.M. Girving eds. "{\it The quantum Hall effect"}, 2nd edition, Springer-Verlag N.Y., 1990.

\bibitem{Barch}  S. Attal, A. Joyce and C.A. Pillet, eds. {\it "Open quantum systems, vol. I, II and III"}, Lectures notes in mathematics, vol.1880-1882, Springer, 2006.

\bibitem{Biane1} Ph. Biane, "Introduction to random walks on non-commuative spaces", in {\it "Quantum potential theory"}, eds. Ph. Biane et al, Springer 2008;

\bibitem{Attal} S. Attal, {\it "Quantum Noise"}, book in preparation, preprint 2010.\

\bibitem{math:dilation} J.L. Sauvageot, "Markov quantum semigroups admit covariant markov C*-dilations", Commun. Math. Phys. 106 (1986) 91-103.

\bibitem{Davies} E.B. Davies, "Markovian master equations, Commun. Math. Phys 39 (1974) 91-110;\\
E.B. Davies, "The classical limit for quantum dynamical semigroups", Commun. Math. Phys. 49 (1976) 113-129.

\bibitem{Markov-limit} J. Lebowitz and H. Spohn, "Irreversible themodynamics for quantum systems weakly coupled to thermal reservoirs", Adv. Chem. Phys. 39 (1978) 109-142.

\bibitem{Huds-parth} R.L. Hudson and K.R. Parthasarathy, "Quantum Ito formula and stochastic evolutions", Commun. Math. Phys. 93 (1984) 301-323.\\
K.R. Parthasarathy, {\it "An introduction to quantum stochastic calculus"}, Monographs in Mathematcis 85, Birkhauser Verlag, 1992.

\bibitem{Lindblad} G. Lindblad, "On the generators of quantum dynamical semigroups", Commun. Math. Phys. 48 (1976) 119-130.

\bibitem{Biane} Ph. Biane, "Ito stochastic calculus and Heisenberg commutation relations", Sto. Proc. Appl. 120 (2010) 698-720.

\end{thebibliography}
\end{document}